\documentclass[hidelinks,nonatbib]{article}
\newcommand{\ifarxiv}[1]{#1}

\ifarxiv{
    \usepackage[final]{nips_2016}
}

\usepackage[utf8]{inputenc} 
\usepackage[T1]{fontenc}    
\usepackage{url}            
\usepackage{booktabs}       
\usepackage{nicefrac}       
\usepackage{microtype}      
\usepackage{enumitem}
\usepackage[usenames, dvipsnames, table]{xcolor}
\usepackage{array, multirow}
\usepackage{amsmath, amssymb, amsfonts}
\usepackage{hyperref}
\usepackage{bbding}
\usepackage{soul}
\usepackage{subfig}

\ifarxiv{
    \usepackage{amsthm}
    \usepackage[pdftex]{graphicx}
    \usepackage[round,comma]{natbib}
}
\usepackage{xspace}
\usepackage[most]{tcolorbox}
\setlist[enumerate]{leftmargin=2em}
\setlist[itemize]{leftmargin=2em}

\newcommand{\xhdr}[1]{\noindent{{\bf #1.}}}

\newcommand{\numberchecked}[1]{{#1}}

\newcommand{\actionitems}[1]{\textbf{Action items:} \begin{itemize} #1 \end{itemize}}
\newcommand{\keyobservations}[1]{\textbf{Key observations:} \begin{itemize} #1 \end{itemize}}
\newcommand{\openquestions}[1]{\textbf{Open problems:} \begin{itemize} #1 \end{itemize}}
\newcommand{\toptwok}{top 2k\xspace}

\newcommand{\tuple}[1]{\{#1\}}

\newcommand{\effectsize}[1]{d= #1}

\newcommand{\incl}[2]{\hfill\includegraphics[width=#1\textwidth]{#2}\hfill} 

\newcommand\blfootnote[1]{%
	\begingroup
	\renewcommand\thefootnote{}\footnote{#1}%
	\addtocounter{footnote}{-1}%
	\endgroup
}
\newcommand{\subsectiontitle}{}
\newcommand{\newsubsection}[1]{\renewcommand{\subsectiontitle}{#1}\subsection{#1}}

\newtcolorbox{Summary}{colback=gray!5!white,
colframe=gray!75!black,fonttitle=\bfseries,top=0.01in,
title=Summary~\arabic{subsection}: \subsectiontitle}

\title{Design and Analysis of the NIPS 2016 Review Process}

\ifarxiv{
    \author{Nihar B. Shah$^*$ \\
     Machine Learning Department, and \\
     Computer Science Department \\
     Carnegie Mellon University \\
     \texttt{\href{mailto:nihars@cs.cmu.edu}{nihars@cs.cmu.edu}} \\
    \And 
    Behzad Tabibian$^*$\thanks{$*$authors contributed equally}\\
     Max Planck Institute for Intelligent Systems, and \\
     Max Planck Institute for Software  Systems \\
     \texttt{\href{mailto:me@btabibian.com}{me@btabibian.com}} \\
    \And 
    Krikamol Muandet \\
     Max Planck Institute for Intelligent Systems \\
     T\"ubingen, Germany \\
     \texttt{\href{mailto:krikamol@tuebingen.mpg.de}{krikamol@tuebingen.mpg.de}} \\
    \AND 
    Isabelle Guyon \\
     Universite Paris-Saclay, France, and \\
     ChaLearn, California \\
     \texttt{\href{mailto:guyon@chalearn.org}{guyon@chalearn.org}} \\
    \And 
    Ulrike von Luxburg \\
     Department of Computer Science \\
     University of Tübingen \\
     \texttt{\href{mailto:luxburg@informatik.uni-tuebingen.de}{luxburg@informatik.uni-tuebingen.de}}
     }
}

\begin{document}
\maketitle

\begin{abstract}
Neural\blfootnote{Nihar Shah, Behzad Tabibian and Krikamol Muandet performed most of the data analysis reported in this paper. Behzad Tabibian and Krikamol Muandet were also the workflow team of NIPS 2016 and were responsible for all the programs,  scripts and CMT-related issues during the review process. Isabelle Guyon and Ulrike von Luxburg were the program chairs of NIPS 2016.} Information Processing Systems (NIPS) is a top-tier annual conference in machine learning. The 2016 edition of the conference comprised more than \numberchecked{2,400} paper submissions, \numberchecked{3,000} reviewers, and 8,000 attendees. This represents a growth of nearly 40\% in terms of submissions, 96\% in terms of reviewers, and over 100\% in terms of attendees as compared to the previous year. The massive scale as well as rapid growth of the conference calls for a thorough quality assessment of the peer-review process and novel means of improvement. In this paper, we analyze several aspects of the data collected during the review process, including an experiment investigating the efficacy of collecting ordinal rankings from reviewers. Our goal is to check the soundness of the review process, and provide insights that may be useful in the design of the review process of subsequent conferences.
\end{abstract}

\section{Introduction}

The review process for NIPS 2016 involved \numberchecked{2,425} papers submitted by \numberchecked{5,756} authors, \numberchecked{100} area chairs, and \numberchecked{3,242} active reviewers submitting \numberchecked{13,674} reviews in total. Designing a review process as fair as possible at this scale was a challenge. In order to scale, all parts of the process have to be as decentralized as possible. Just to get a feeling, if the two program chairs were supposed to take final decisions just for the 5\% most  challenging submissions, which means that they would have to read and decide on 150 papers --- this is the scale of a whole conference such as COLT. Furthermore, the complexity of the logistics and software to manage the review process is rather high already. A controlled experiment~\citep{nips14experiment} from NIPS 2014 has shown that there is a high disagreement in the reviews. Hence the primary goal must be to keep bias and variance of the decisions as small as possible.

In this paper, we present an analysis of many aspects of the data collected throughout the review phase of the NIPS 2016 conference, performed subsequent to the completion of the review process. Our goal in this analysis is to examine various aspects of the data collected from the peer review process to check for any systematic issues. Before delving into the details, the reader should importantly note the following limitations of this analysis: 
\begin{itemize}
   \item There is no ground truth ranking of the papers or knowledge of the set of papers which should ideally have been accepted.
   \item The analysis is post hoc, unlike the controlled experiment from NIPS 2014~\citep{nips14experiment}.
   \item The analysis primarily evaluates the ratings and rankings provided by reviewers, and does not study the textual comments provided by the reviewers.
\end{itemize}
The analysis is used to obtain insights into the peer-review process, usable suggestions for subsequent conferences, and important open problems towards improving peer-review in academia.

Here is a summary of our findings:
\begin{enumerate}[label=(\roman*)]
\item there are very few positive bids by reviewers and area chairs (Section~\ref{SecBids}),
\item graph-theoretic techniques can be used to ensure a good reviewer assignment (Section~\ref{SecGraph}),
\item there is significant miscalibration with respect to the rating scale (Section~\ref{SecScoreDistribution}),
\item review scores provided by invited and volunteer reviewers have comparable biases and variance; junior reviewers report a lower confidence (Section~\ref{SecDifferentReviewers}),
\item there is little change in reviewer scores after rebuttals (Section~\ref{SecDiscRebuttals}),
\item there is no observable bias towards any research area in accepted papers (Section~\ref{SecAreas}),
\item there is lower disagreement among reviewers in NIPS 2016 as compared to NIPS 2015 (Section~\ref{SecRandomness}),
\item significant fraction of scores provided by the reviewers are tied and ordinal rankings can ameliorate this issue (Section~\ref{SecOrdinal}),
\item there are some inconsistencies in the reviews and these can be identified in an automated manner using ordinal rankings (Section~\ref{SecAnomalies}).
\end{enumerate}

We describe the review procedure followed at NIPS 2016 in Section~\ref{SecProcedure}. We present an elaborate description of the analysis and the results in Section~\ref{SecAnalysis}. Alongside each analysis, we present a set of key observations, action items for future conferences, and some open problems that arise out of the analysis. We conclude the paper with a discussion in Section~\ref{SecConclusion}.


\section{Review procedure}
\label{SecProcedure}

In this section, we present an overview of the design of the review process at NIPS 2016.

\subsection{Selecting area chairs and reviewers}

{\bf Area Chairs (ACs)} are the backbone of the NIPS reviewing process. Their role is similar to that of an associate editor for a journal. Each AC typically handles 20-30 submissions, so with an estimated number of submissions between 2000 and 3000, we needed to recruit about 100 area chairs. As it is impossible to intimately know all the diverse research areas covered by NIPS, we came up with the following procedure. 
We asked the NIPS Board and all the ACs of NIPS from the past two years to nominate potential ACs for this year. In this manner, we covered the entire variety of NIPS topics and obtained qualified suggestions. We obtained around 350 suggestions.
We asked the NIPS Board to go through the list of suggested ACs and vote in favor of suggested ACs.
 We also accounted for the distribution of subject areas of submitted papers of the previous year's NIPS conference.
 Combining all these inputs, we compiled a final list of ACs: by the end of January we had recruited exactly 100 ACs. 
   In a subsequent step, we formed ``buddy pairs'' among the ACs. Based on the ACs preferences, each AC got assigned a buddy AC. We revisit the role of buddy pairs in more detail later.

 The process of {\bf recruiting reviewers} is time consuming, it essentially went on from January until the submission deadline at end of May. A significant departure from the review processes of NIPS from earlier years, this time we had two kinds of reviewers, ``invited senior reviewers'' (Pool 1) and ``volunteer reviewers'' (Pool 2):
\begin{itemize}
\item  {\bf Pool 1,  invited senior reviewers:}  We asked all ACs to suggest at least 30 reviewers who have completed their PhDs (however, this requirement was not strictly observed by all ACs). We obtained 2500 suggested experienced reviewers. We invited all of them, and 1100 accepted. We then asked all confirmed reviewers to ``clone themselves'' by inviting at least one researcher with a similar research background and with at least as good a qualification as themselves. This resulted in an additional 500 experienced reviewers.

\item {\bf Pool 2, volunteer author-reviewers:} The rapid growth in the number of submissions at NIPS poses the formidable challenge of accordingly scaling the number of reviewers. An obvious mean to achieve this objective is to ask authors to become reviewers as well. This idea has been used in the past, for example, to evaluate NSF grant proposals \citep{Mervis14} or to allocate telescope time \citep{MerSar09}. In order to implement this idea, without constraining unwilling authors, we requested authors to volunteer during the submission process by naming at least one author per paper as volunteer reviewers. We invited all of them and about 2000 of the volunteers accepted the invitation. 
\end{itemize} 

The area chairs were aware of the respective pools to which each of their reviewers belonged.
 The number of reviewers that we eventually ended up with are as follows: 
\begin{center}
 \begin{tabular}{|p{4.78cm}|>{\centering\arraybackslash}p{2.72cm}|>{\centering\arraybackslash}p{2.72cm}|>{\centering\arraybackslash}p{1.5cm}|}
 \hline
 & Senior researchers / faculty & Junior researchers / postdocs & PhD students \\
 \hline
Pool 1: Invited reviewers &  1236 & 566 &  255 \\
Pool 2: Volunteer reviewers & 143 & 206 & 827 \\
\hline
\end{tabular}
\end{center}


\subsection{Assignment of papers to reviewers and area chairs}

The assignment of papers to area chairs was made in the following manner. Prior to the review process, the ACs (and reviewers) were allowed to see the list of submitted papers and ``bid'' whether they were interested or disinterested in handling (or reviewing) any paper. For any paper, an AC (or reviewer) could either indicate ``Not Willing'' or ``In-a-pinch'' -- which we count as negative bids, or indicate ``Willing'' or ``Eager'' -- which we count as positive bids, or choose to not bid for that paper. The Toronto paper matching system or TPMS was then employed to compute an affinity score for every AC (and reviewer) with every submitted paper based on the content of the paper and the academic profile of the AC or reviewer. In addition, every AC (and reviewer) as well as the submitter of every paper was asked to select a set of most relevant subject areas, and these subject areas were also employed to compute a similarity between each AC (and reviewer) and paper.

Based on the similarity scores and bids, an overall similarity score is computed for every \tuple{paper, AC} and every \tuple{paper, reviewer} pair: $\text{score} = b (  s_{\text{affinity}}+ s_{\text{subject}})$, where $s_{\text{affinity}} \in [0,1]$ is the affinity score  obtained from TPMS, $s_{\text{subject}} \in [0,1]$ is the score obtained by comparing the subject areas of the paper and the subject areas selected by the AC or reviewer, and $b \in [0.25,1]$ is the bidding score provided by the AC or reviewer. Based on these overall similarity scores, a preliminary paper assignment to ACs was then produced in an automated manner using the TPMS assignment algorithm~\citep{charlin2013toronto}. The ACs were given a provision to decline handling certain papers for various reasons such as  conflicts of interest. These papers were re-assigned manually by the program chairs.

The AC of each paper was responsible to first assign one senior, highly qualified reviewer manually. Two more invited reviewers from pool 1 and three volunteer reviewers from pool 2 were then assigned automatically to each paper using the same procedure as described above. The ACs were asked to verify whether each of their assigned papers had at least 3 highly competent reviewers; the ACs could manually change reviewer assignments to ensure that this is the case.  During the decision process, additional emergency reviewers were invited to provide complementary reviews if some of the reviewers had defected or if  no consensus was reached among the selected reviewers.


\subsection{Review criteria and scores}
\label{SecReviewCriteria}

We completely changed the scoring method this year. In previous years, NIPS papers were rated using a {single score} between 1 and 10. A single score alone did not allow reviewers to give a differentiated quantitative appreciation on various aspect of paper quality. Furthermore, the role of the ACs was implicitly to combine the decisions of the reviewers (late integration) rather than combining the reviews to make the final decision (early integration). Introducing multiple scores allowed us to better separate the roles: the reviewers were in charge of evaluating the papers; the ACs were in charge of making decisions based on all the evaluations. Furthermore the multiple specialized scores allowed the ACs to guide reviewers to focus discussions on ``facts'' rather than ``opinion'' in the discussion phase. We asked reviewers to provide a separate score for each of the following four features: 
\begin{itemize}[topsep=-1pt]
\itemsep0em
\item    Technical quality, 
\item    Novelty/originality, 
\item    Potential impact or usefulness, 
\item    Clarity and presentation.
\end{itemize}
The scores were on a scale of 1 to 5, with the following rubric provided to the reviewers:  
\begin{itemize}[topsep=-1pt]
\itemsep0em
\item[] 5  = Award level (1/1000 submissions), 
\item[]   4 = Oral level (top 3\% submissions), 
\item[]    3 = Poster level (top 30\% submissions),
\item[]    2 = Sub-standard for NIPS, 
\item[]    1 =  Low or very low. 
\end{itemize}
The scoring guidelines also reflect the hierarchy of the papers: the conference selects the top few papers for awards, the next best accepted papers are presented as oral presentations, and the remaining accepted papers are presented as posters at the conference. The scores provided by reviewers had to be complemented by justifications in designated text boxes.  We also asked the reviewers to flag ``fatal flaws'' in the papers they reviewed. For each paper, we also asked the reviewers to declare their overall ``level of confidence'': 
\begin{itemize}[topsep=-1pt]
\itemsep0em
\item[] 3 = Expert (read the paper in detail, know the area, quite certain of opinion),
\item[] 2 = Confident (read it all, understood it all reasonably well),
\item[] 1 = Less confident (might not have understood significant parts).
\end{itemize}


\subsection{Discussions and rebuttals}

Once most reviews were in, authors had the opportunity to look at the reviews and write a rebuttal. One section of the rebuttal was revealed to all the reviewers of the paper, and a second section was private and visible only to the ACs. Some reviews were still missing at this point, but it would not have helped to delay the rebuttal deadline as the missing reviews trickled in only slowly. Subsequently, ACs and reviewers engaged in discussions about the pros and cons of the submitted papers. To support the ACs, we sent individual reports to all area chairs to flag papers whose reviews were of too low confidence, too high variance or where reviews were still missing. In many cases, area chairs recruited additional emergency reviewers to increase the overall quality of the decisions.


\subsection{Decision procedure} 

The decision procedure involved making an acceptance or rejection decision for each paper, and furthermore, to select a subset of (the best) accepted papers for oral presentation.

We introduced a decentralized decision process based on pairs of ACs (``buddy pairs''). Each AC got assigned one buddy AC. Each pair of buddy ACs was responsible for all papers in their joint bag and made the accept/reject decisions jointly, following guidelines given by the program chairs. Difficult cases were taken to the program chairs, which included cases involving conflicts of interest and plagiarism. In order to harmonize decisions across buddy pairs, all area chairs had access to various statistics and histograms over the set of their papers and the set of all submitted papers. To decide which accepted paper would get an oral presentation, each buddy pair was asked to champion one or two papers from their joint bag as a candidate for an oral presentation. The final selection was then made by the program chairs, with the goals of exhibiting the diversity of NIPS papers and exposing the community with novel and thought provoking ideas. In the end,  \numberchecked{568} papers got accepted to the conference, and 45 of these papers were selected for oral presentations.

Like previous years, we adopted a ``double blind'' review policy. That is, the author(s) of each paper did not get to know the identity of the reviewers and vice versa throughout the review process. ACs got to know the identity of the reviewers and the author(s) for the papers under their responsibility. During the discussion phase, reviewers who reviewed the same papers got to know each others' identity. Lastly, PCs and program managers had access to all information about the submissions, the ACs, the reviewers, and the authors.


\subsection{Experimental ordinal reviews}

In the main NIPS 2016 review process, we elicited only cardinal scores from the reviewers -- one score in 1 to 5 for each of four features. Subsequent to the review process, we then requested each reviewer to also provide a total ranking of the papers that they reviewed. We received   rankings from a total of 2189 reviewers. Note that the collection of ordinal data was performed subsequent to the normal review submission but before release of the final decisions. The ordinal data was not used as a part of the decision procedure in the conference.


\section{Detailed analysis}\label{SecAnalysis}

In this section, we present details of our analyses of the review data and the associated results. Each subsection contains one analysis and concludes with a summary that highlights the key observations, concrete action items for future conferences, and open problems that arise from the analysis. 

The results are computed for a snapshot of reviews at the end of the review process when the acceptance decisions were made. This choice does not affect the results since there was very little change in the scores provided by reviewers across different time instants. All t-tests conducted correspond to two-sample t-tests with unequal variances. All mentions of p-values correspond to two-sided tail probabilities. All mentions of statistical significance correspond to a p-value threshold of 0.01 (we also provide the exact p-values alongside). Multiple testing is accounted for using the Bonferroni correction. The effect sizes refer to Cohen's d. Wherever applicable, the error bars in the figures represent 95\% confidence intervals. 

Wherever applicable, we also perform our analyses on a subset of the submitted papers which we term as the \toptwok papers. The \toptwok papers comprise all of the 568 accepted papers, and an equal number (568) of the rejected papers. The 568 rejected papers are chosen as those with maximum scores averaged across all reviewers and all features.


\newsubsection{Reviewer bids}
\label{SecBids}

A large number of conferences in computer science ask area chairs and/or reviewers to bid which papers they would like or not like to review, in order to obtain a better understanding of the expertise and the preferences of reviewers. Such an improved understanding is desirable as it leads to a more informed assignment of reviewers to papers, thereby improving the overall quality of the review process.

\begin{figure}[t!]
\ifarxiv{\setlength{\belowcaptionskip}{-10pt}}
\centering
\begin{tabular}{c c c}
\incl{0.45}{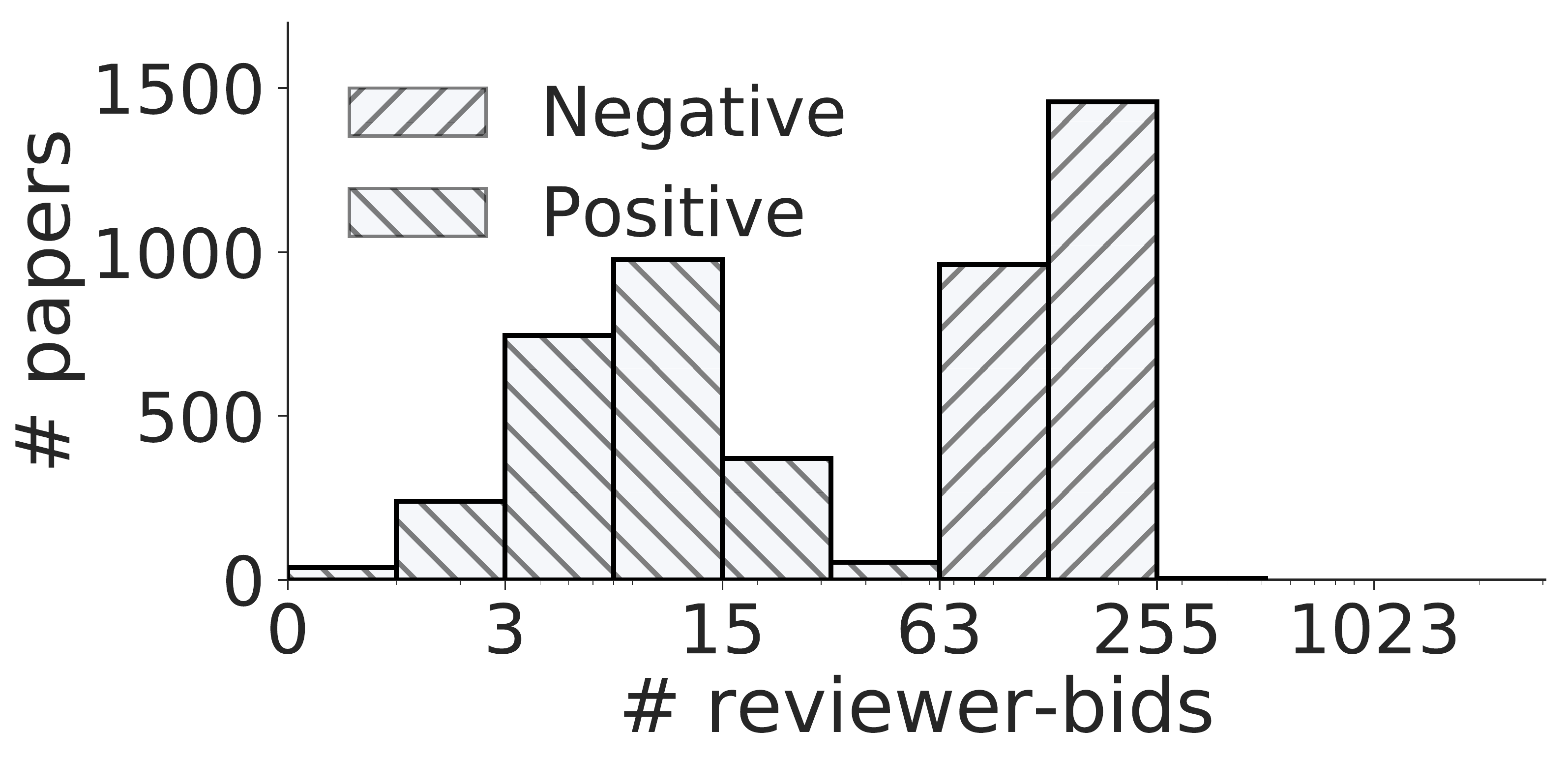} &
\quad &
\incl{0.45}{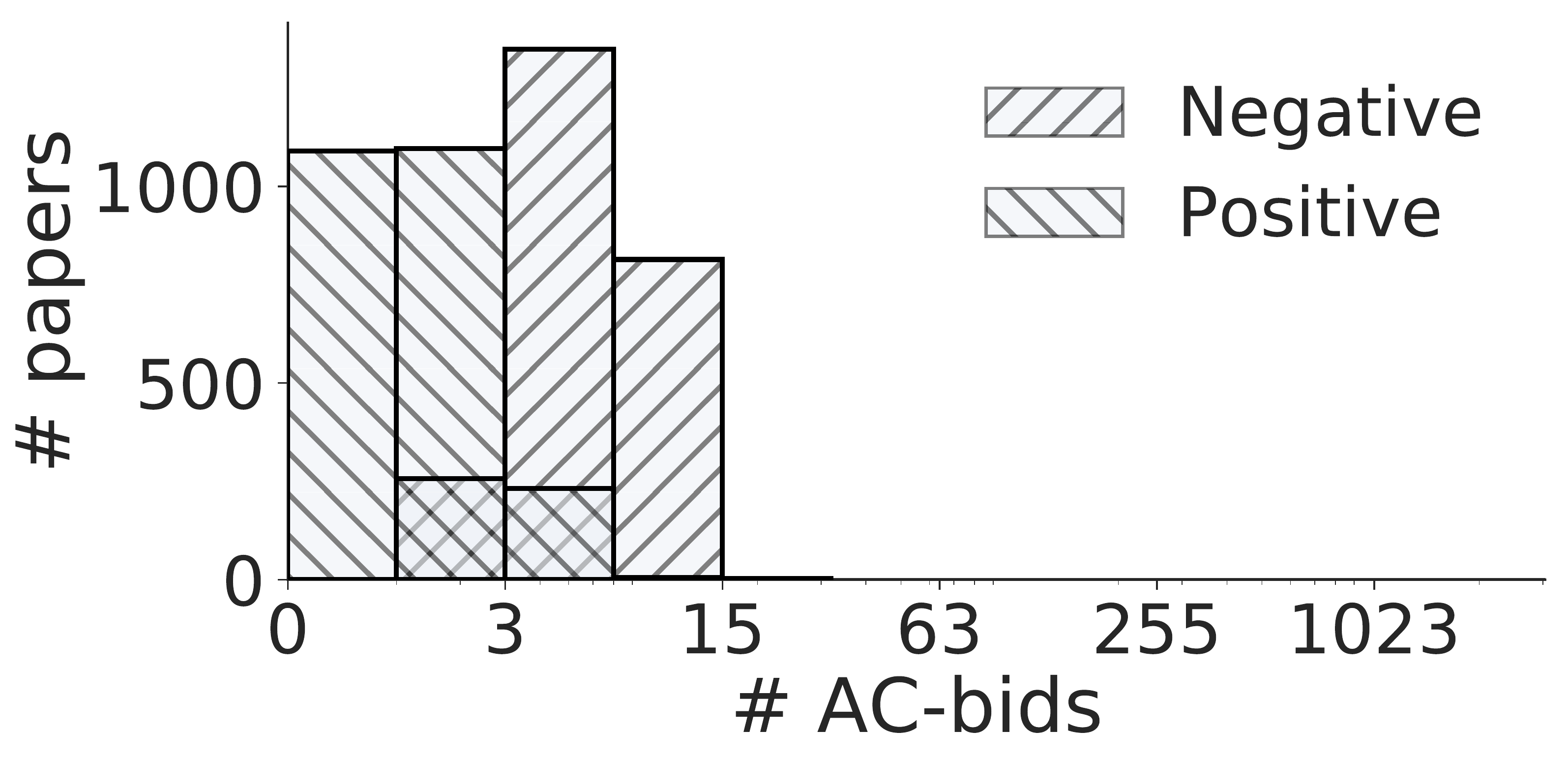} \\ (a) total number of bids by reviewers per paper & \quad & (b) total number of bids by ACs per paper\\~\\
\incl{0.45}{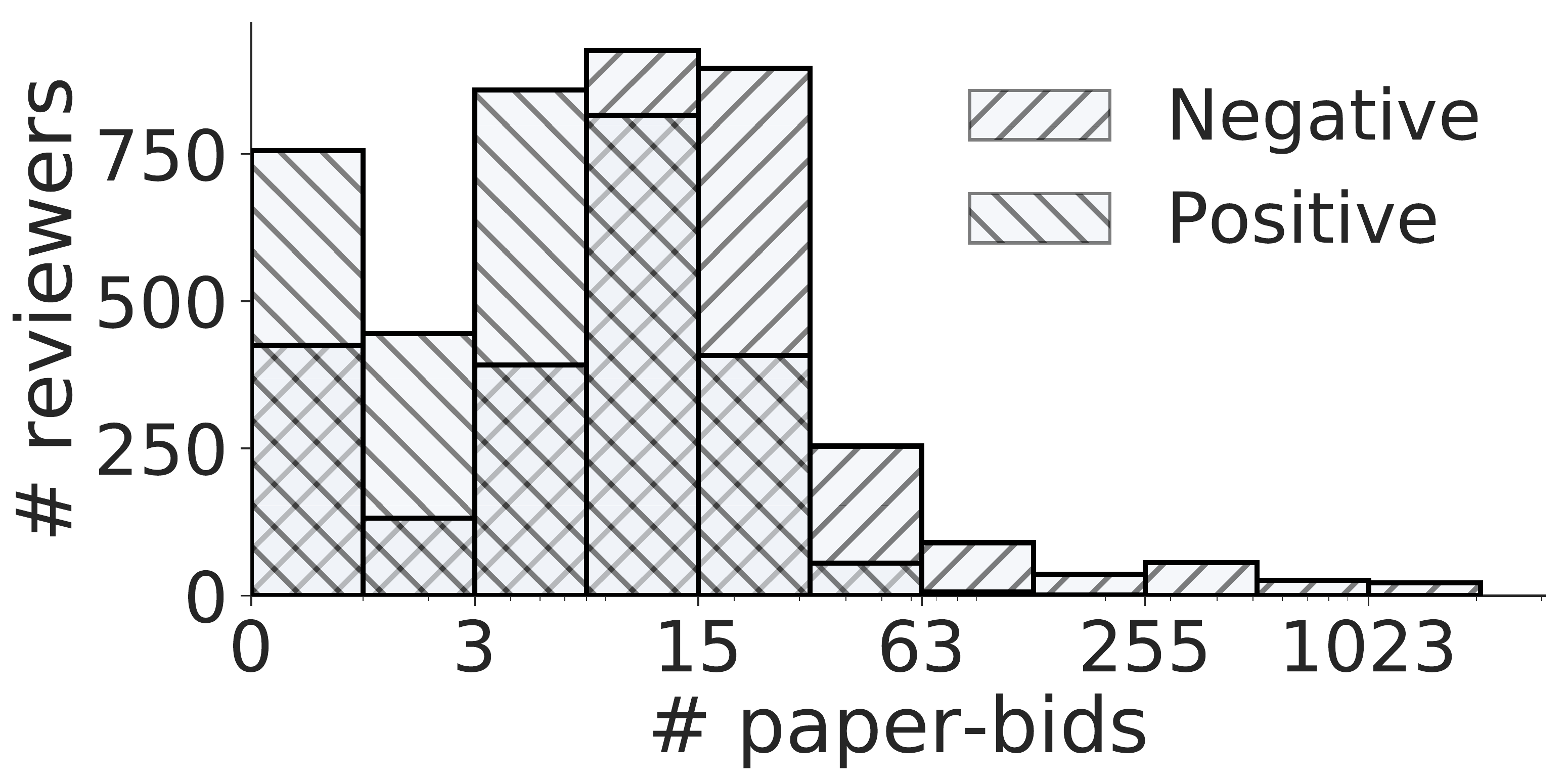} &
\quad &
\incl{0.45}{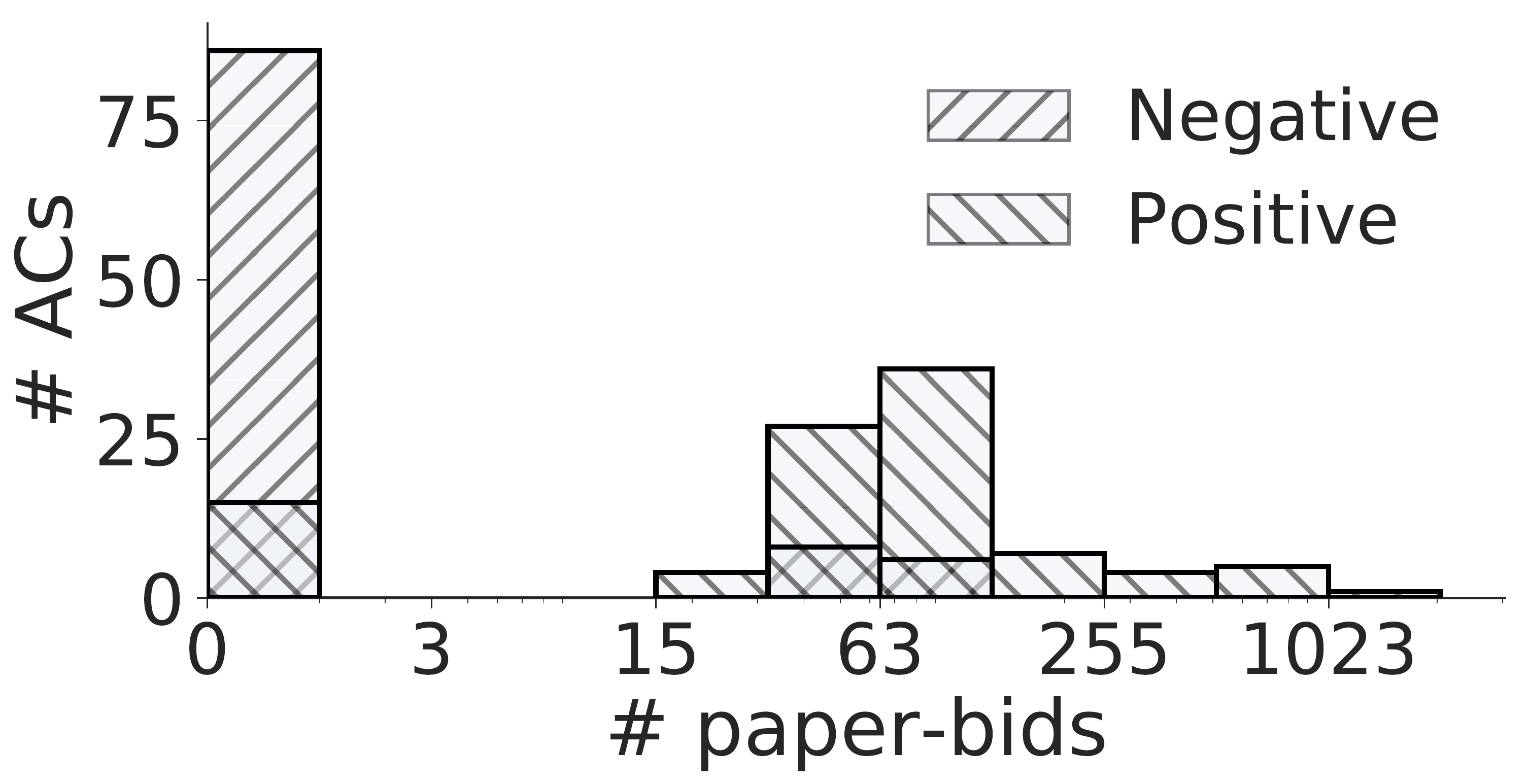}\\
(c) number of bids per reviewer  & \quad & (d) number of bids per AC 
\end{tabular}  
\caption{Histogram of number of positive and negative bids (x-axis; on a logarithmic scale) per entity (counts on y-axis) for various entities. The ``not willing'' and ``in-a-pinch'' bids were considered negative bids, whereas ``willing'' and ``eager'' bids were considered positive bids. The first column in each histogram represents number of entities with 0 bids. For example, the first column of panel (c) depicts that 756 reviewers made zero positive bids and 425 reviewers made zero negative bids.
\label{fig:bids}}
\end{figure}

Figure~\ref{fig:bids} depicts the distribution of number of bids on papers submitted by area chairs and reviewers in NIPS 2016. Panels (a) and (b) of the figure depict the distribution of counts per paper for reviewers and area chairs respectively; panels (c) and (d) depict the distribution per area chairs and reviewers. From the data, we observe that there are very few positive bids, but a considerably higher number of negative bids. 

The distribution of number of bids by reviewers is skewed by few reviewers who bid (positive and negative) on too many papers: \numberchecked{27\%} of reviewers make \numberchecked{90\%} of all bids, and \numberchecked{50\%} of reviewers make \numberchecked{90\%} of all positive bids. Moreover, there are \numberchecked{148} reviewers with no (positive or negative) bids and \numberchecked{1201} reviewers with at most 2 positive bids. In comparison, NIPS 2016 assigned at least 3 papers to most reviewers and many conferences do likewise. We thus observe that a large number of reviewers do not even provide positive bids amounting to the number of papers they would review. As a consequence of the low number of bids by reviewers, we are left with \numberchecked{278} papers with at most 2 positive bids and \numberchecked{816} papers with at most 5 positive bids. In contrast, NIPS 2016 assigned 6 reviewers to most papers. There is thus a significant fraction of papers with fewer positive bids than the number of requisite reviewers. Finally there are \numberchecked{1090} papers with no positive bids by any AC.

\begin{Summary}

\keyobservations{\item There are very few positive bids, with \numberchecked{278} papers receiving at most 2 positive bids and \numberchecked{816} papers receiving at most 5 positive bids.

\item From the reviewers' side, the bids are highly skewed: \numberchecked{50\%} of reviewers make \numberchecked{90\%} of all positive bids, \numberchecked{148} reviewers make no (positive or negative) bids, and \numberchecked{1201} reviewers make at most 2 positive bids.}

\actionitems{\item When a reviewer logs into the system, the interface can show the unbid papers on top.

\item Inform reviewers of the procedure employed to use their bids for assigning papers. Make reviewers aware of the benefits of bidding, such as receiving more relevant papers to read and serving the community by improving the review process.
}

\openquestions{\item How to incentivize more (positive) bids so that the organizers understand preferences better for accurate reviewer assignment? 

\item Design a principled means of combining bids, paper content-reviewer profile similarity, and subject similarity.

}
\end{Summary}


\newsubsection{Reviewer assignment}
\label{SecGraph}

\begin{figure}[t!]
\ifarxiv{\setlength{\belowcaptionskip}{-10pt}}
\centering
\begin{tabular}{c c c}
\incl{0.35}{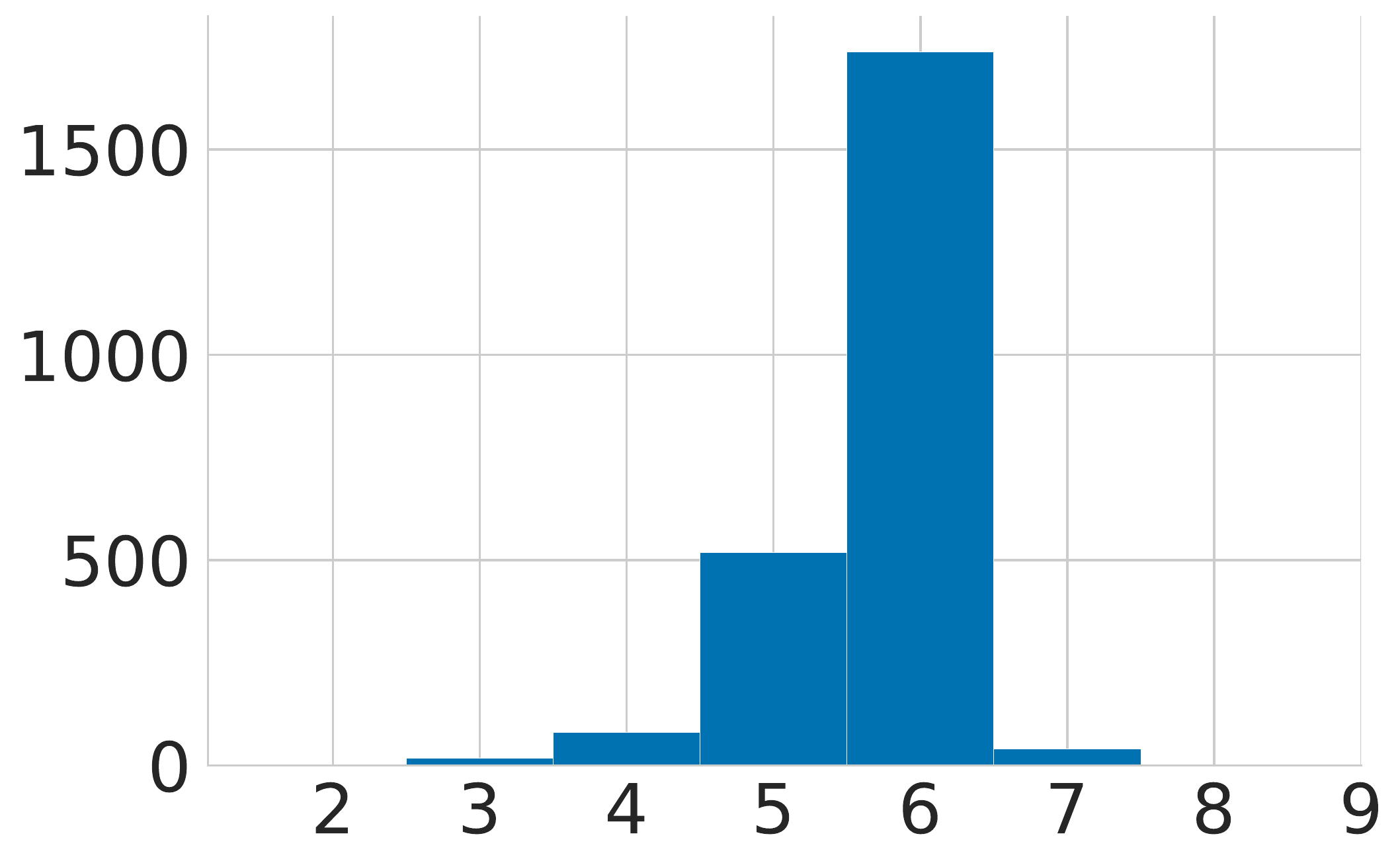} &
\qquad &
\incl{0.35}{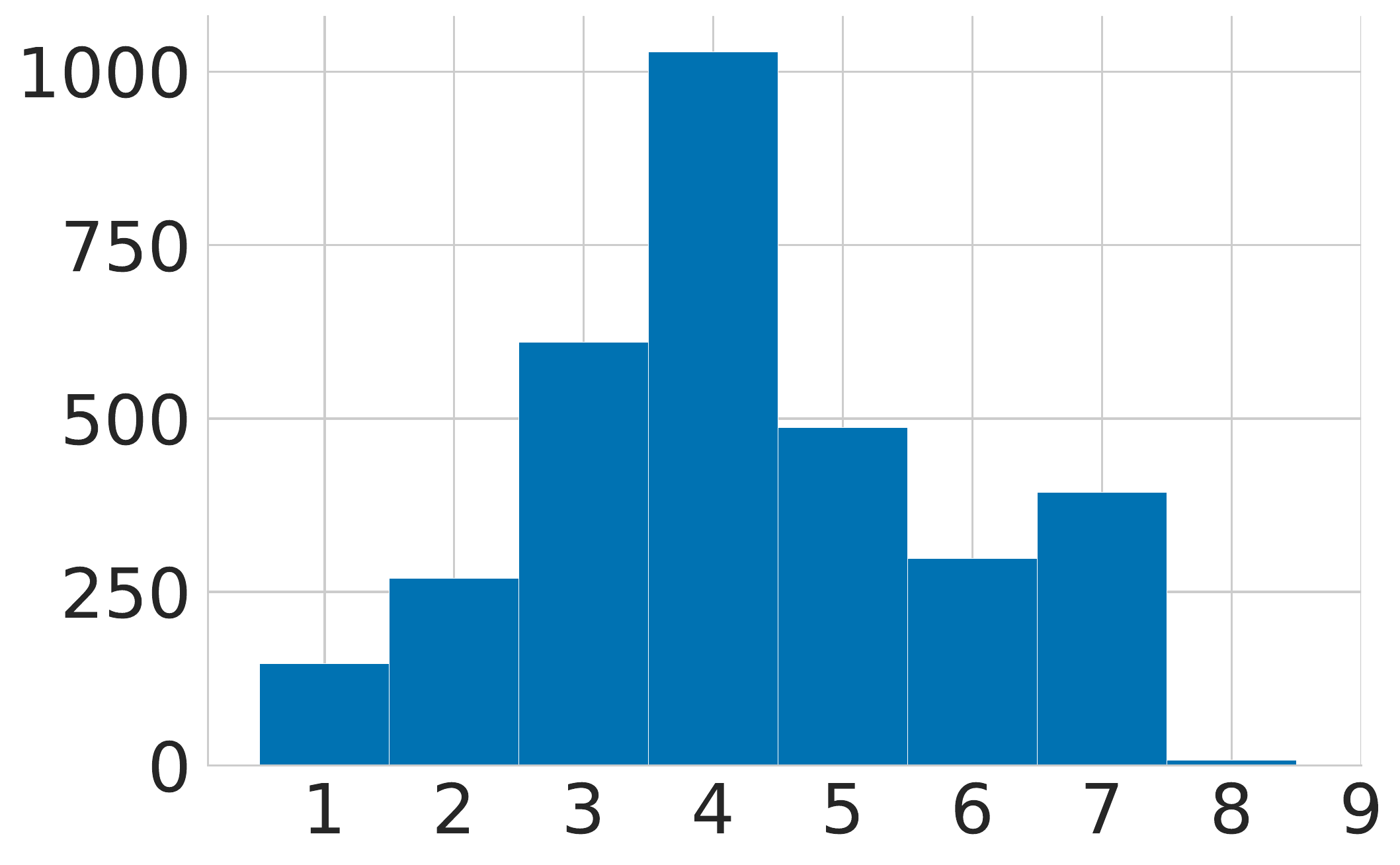} \\
(a) number of reviewers per paper & \qquad & (b) number of papers per reviewer
\end{tabular}  
\caption{Histogram of number of reviews.
\label{fig:counts}}
\end{figure}

Figure~\ref{fig:counts} depicts the histograms of the  number of reviewers assigned per paper, and the number of papers handled by each reviewer.

\begin{figure}[b!]
\centering
\begin{tabular}{c c c}
\incl{0.37}{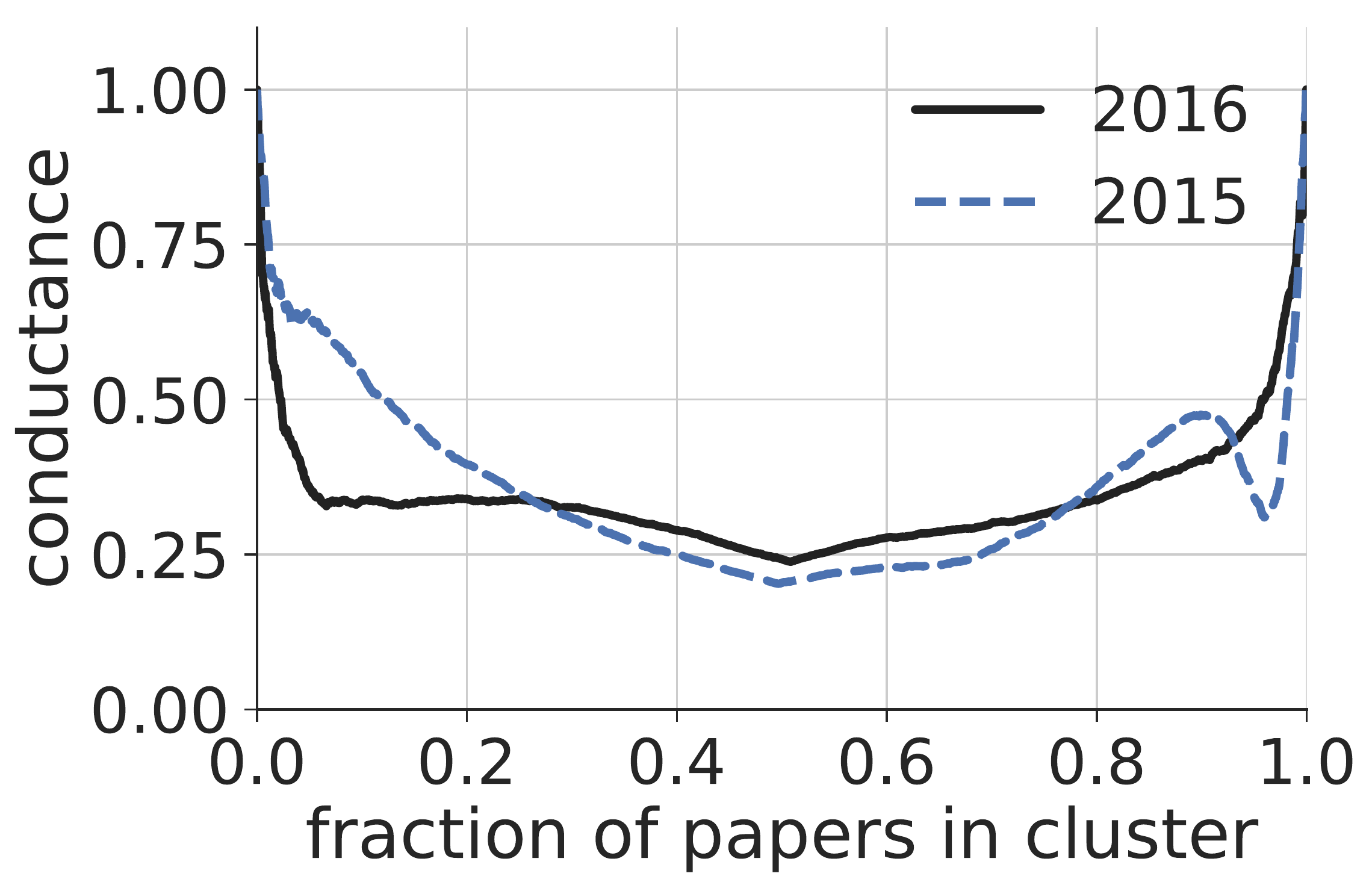} &
\quad &
\incl{0.37}{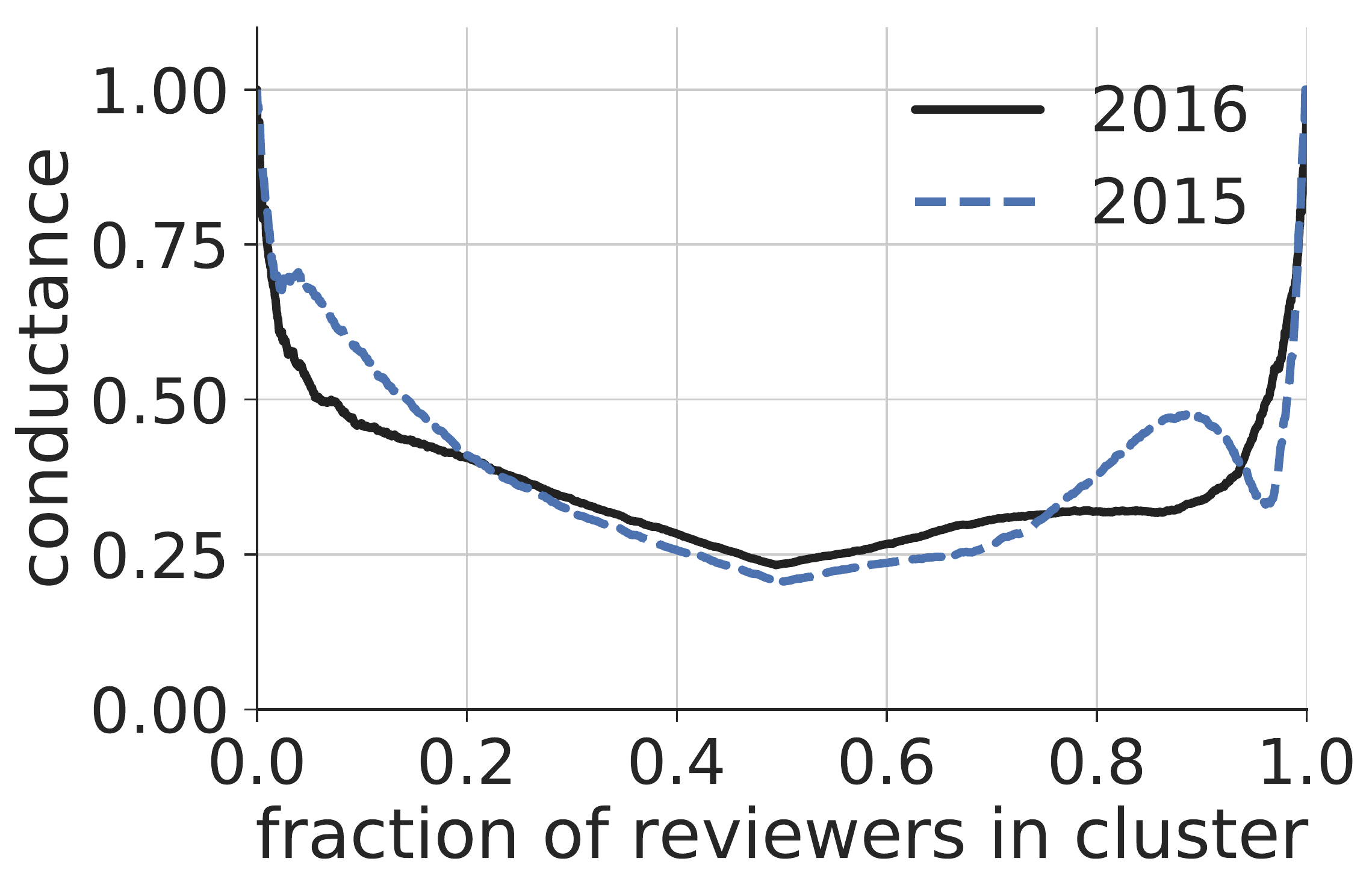}\\
(a) paper graph  & 
\quad &
(b) reviewer graph\\
\end{tabular}
\caption{Conductance value as function of varying cluster size. The x-axes in these plots is the normalized cluster size $k/\lvert V\rvert$. \label{FigConductanceGraph}}
\end{figure}

In order to ensure that the information about each paper ``spreads'' across the entire system, it is important that there is no set of reviewers or papers that has only a small overlap with the remaining reviewers and papers~\citep{olfati2007consensus,shah2016estimation}. 
To analyze whether this was the case, we considered two graphs. We built a {\it reviewer graph} that has reviewers as vertices, and an edge between any two reviewers if there exists at least one paper that has been reviewed by both of them. Analogously we built a {\it paper graph}, where vertices represent papers, and we connect two papers by an edge if there exists a reviewer who has reviewed both papers. Note that the graph structure is in part dictated by a constraint on the maximum number of papers per reviewer, as well as the specified number of reviewers per paper.

Our objective is to examine the structure of the graphs and determine if there were any separated communities of nodes. In order to do so, we employ a method based on spectral clustering. Formally, denote any graph as $G=(V,E)$ where $V$ is set of nodes, and $E$ is the set of (undirected) edges between nodes, and $\lvert V\rvert$ is number of nodes in the graph. We can denote graph connectivity by its associated adjacency matrix $A$ which is a $(\lvert V\rvert \times \lvert V\rvert)$ matrix; we have $A_{ij}=1$ if there is an edge between nodes $i$ and $j$ and $A_{ij}=0$ otherwise. With this notation, a quantity known as the ``conductance'' $\Phi$ of any set of nodes $S \subset V$ is then defined as:
\begin{figure}[t!]
\setlength{\abovecaptionskip}{4pt}
\centering
{
\setlength{\tabcolsep}{0pt}
\begin{tabular}{c c c c}
\multicolumn{2}{c}{\textbf{Reviewers}}&\multicolumn{2}{c}{\textbf{Papers}}\\
\incl{0.25}{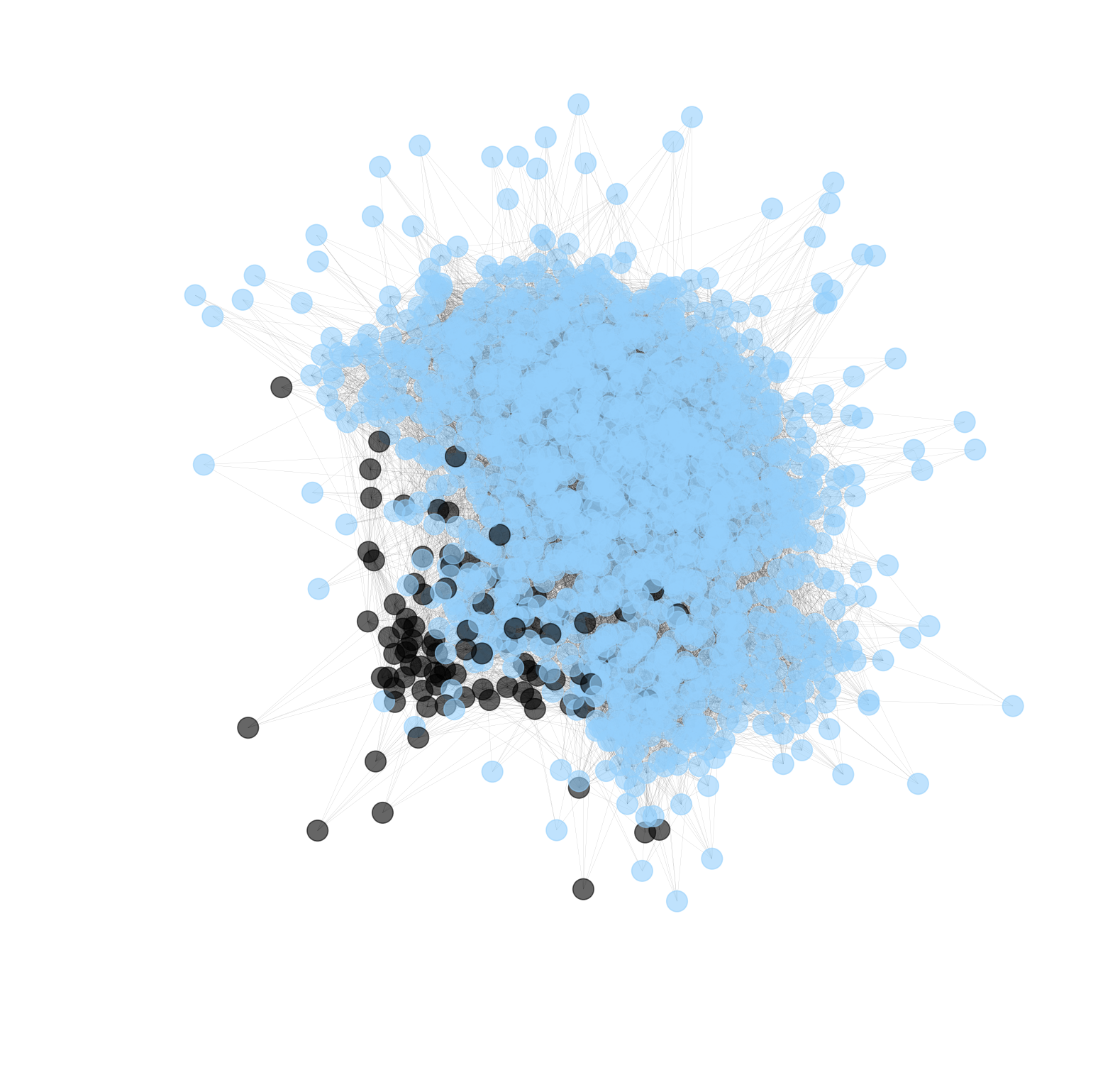}&
\incl{0.25}{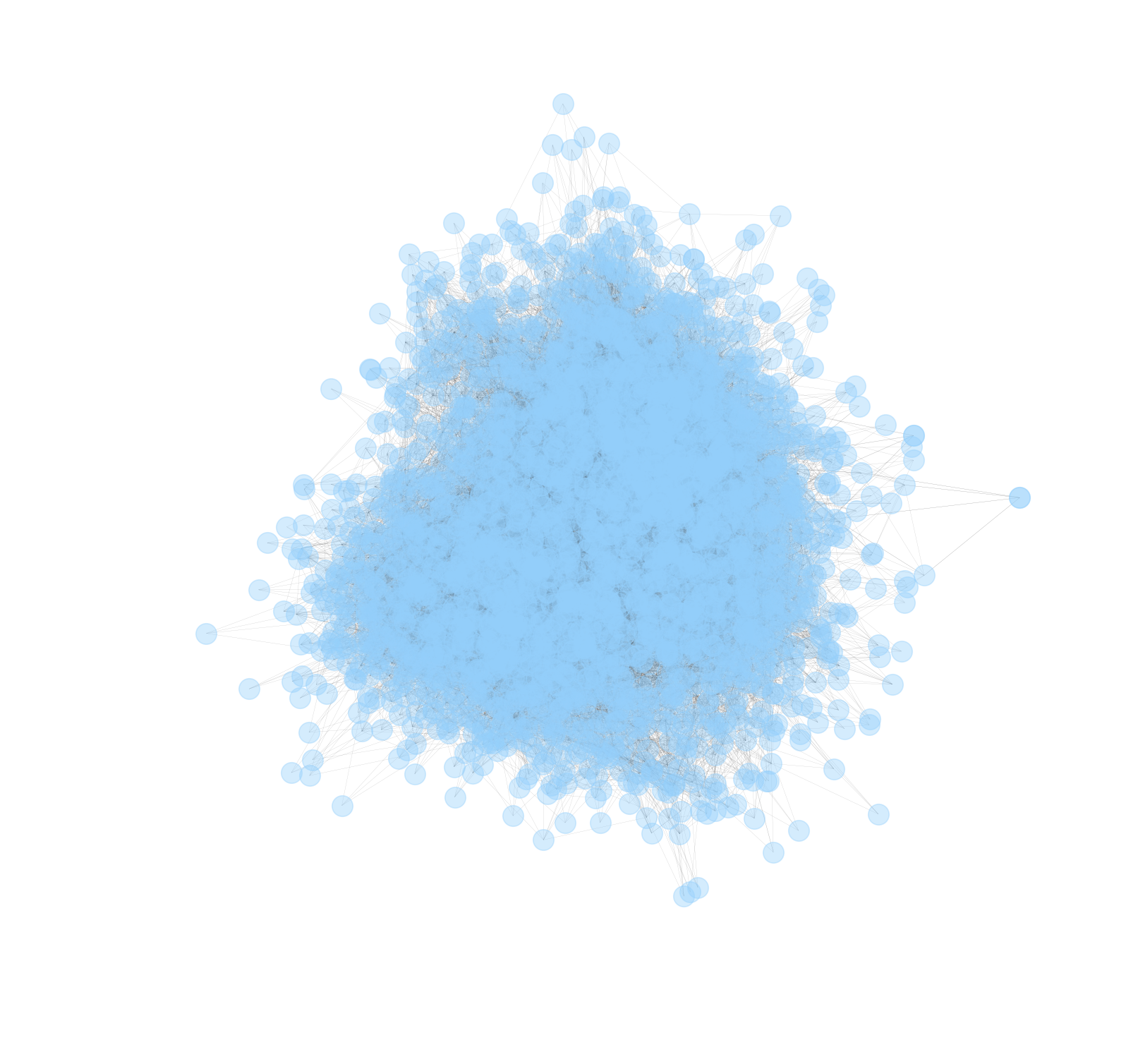}&
\incl{0.25}{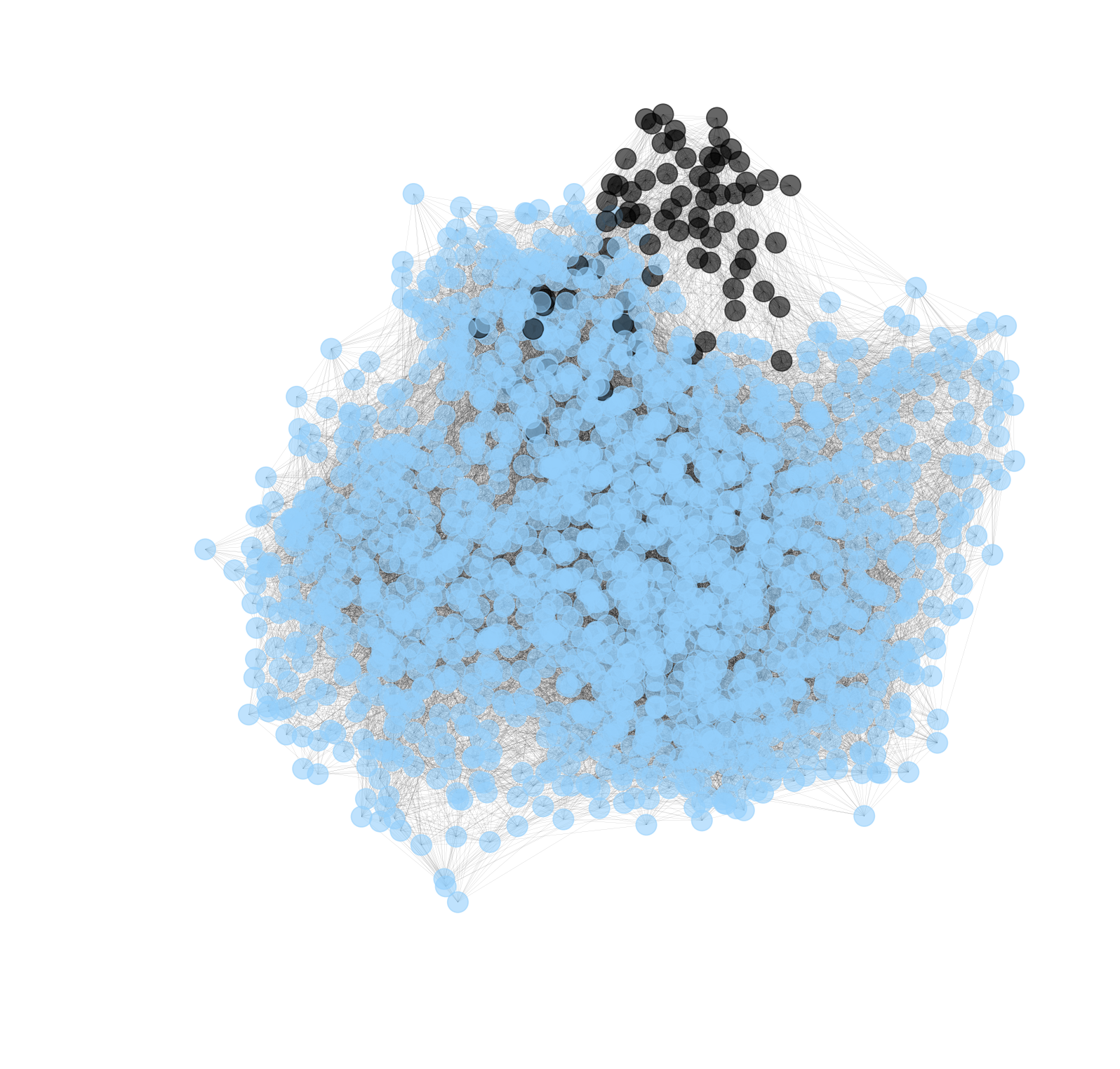}&
\incl{0.25}{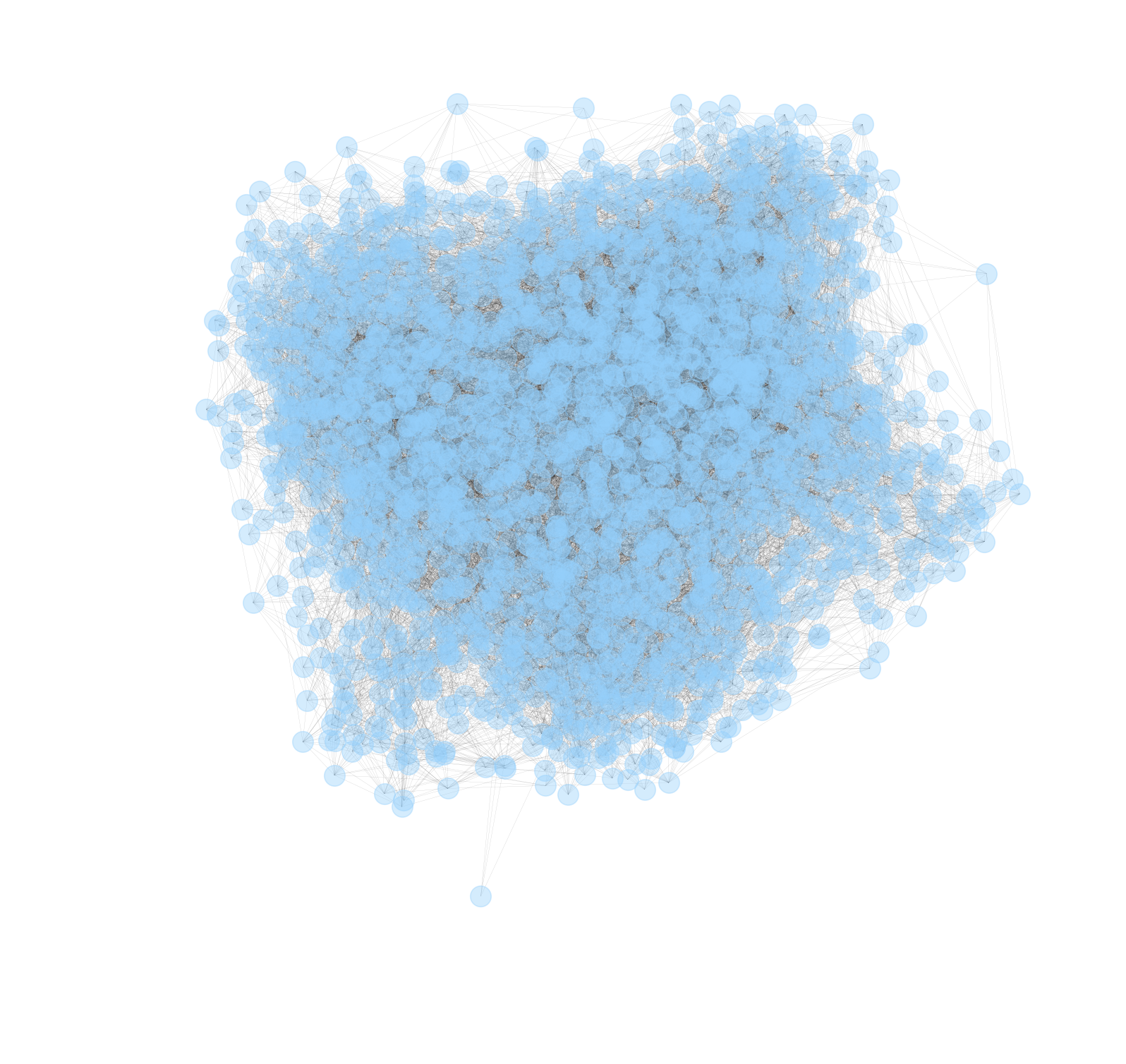}\vspace{-.1in}\\
NIPS 2015& NIPS 2016&NIPS 2015&NIPS 2016\\
\end{tabular}
}
\caption{Graphs depicting connectivity of reviewers and that of papers for NIPS 2015 and NIPS 2016. The nodes in black (dark) show set of nodes identified by the local minima in the conductance plots (Figure~\ref{FigConductanceGraph}) for NIPS 2015, and the remaining nodes are plotted in blue (light). \label{FigureReviewerGraph}}
\end{figure}
\begin{figure}[t!]
\setlength{\belowcaptionskip}{-20pt}
\setlength{\abovecaptionskip}{0pt}
\centering
\incl{1}{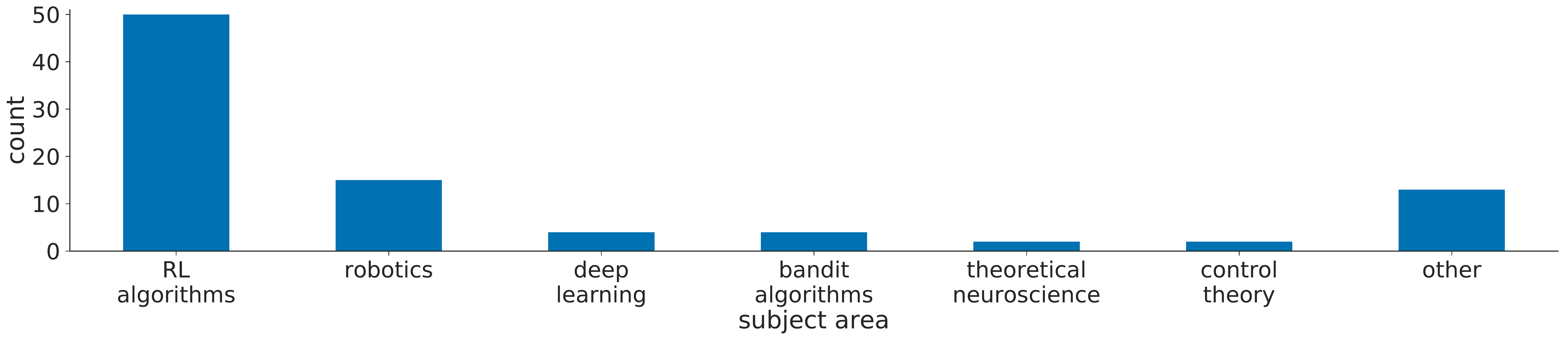}
\caption{Histogram of subject areas in the identified cluster (from Figure~\ref{FigureReviewerGraph}) of reviewers in NIPS 2015 which is not well connected with the set of remaining reviewers.\label{FigureClusterSubjects}}
\end{figure}
\begin{align*}
    \Phi(S) = \frac{\sum \limits_{i\in S, j\not\in S}{A_{ij}}}{\max\{\lvert S \rvert,\lvert V \backslash S \rvert \}},
\end{align*}
where $V \backslash S$ is the complement of set $S$. A lower value of the conductance indicates that the nodes in the cut are less connected to the remaining graph. 
Next, with a minor abuse of notation, the conductance of a graph as function of cluster sizes is defined as:
\begin{align*}
    \Phi(k) = \min \limits_{S \in V, \lvert S \rvert=k} \Phi(S),
\end{align*}
for every $k \in \{1,\ldots,\lvert V\rvert-1\}$. The plot of $k$ versus $\Phi(k)$ is called a Network Community Profile or NCP plot~\citep{leskovec2008statistical}. The NCP plot measures the quality of the least connected community (lowest conductance) in a large network, as a function of the size of the community. Although computing the function $\Phi(k)$ exactly may be computationally hard, an approximate value can be computed using a simple ``second left eigenvector'' procedure (Section 2.3 of \citealp{benson2015tensor}). A well connected graph would have a smooth plot of $\Phi(k)$ with a minima at around $k = \lvert V\rvert/2$.

Figure~\ref{FigConductanceGraph} shows the NCP plot for an increasing number of papers (respectively reviewers) in the paper graph (respectively reviewer graph). For reference we also plot the same curve for graphs associated with NIPS 2015 conference. Both plots for NIPS 2015 have local minima at around $k = 0.96 \lvert V\rvert$, indicating that there is a densely connected community of reviewers and papers that are not well connected with the rest of the graph. In contrast, the plot associated with NIPS 2016 decreases smoothly and reaches its global minimum when half of the nodes are in one cluster and the other half in another cluster, indicating an absence of such a fragmentation.

In Figure~\ref{FigureReviewerGraph}, we plot the graph of reviewers and papers using the algorithm of~\citet{fruchterman1991graph}. In these figures we identify the set of nodes that are identified using the aforementioned NCP method; these nodes are colored black (dark) in the figure in contrast to the blue (light) color of the remaining nodes. We can see from the Figure~\ref{FigureReviewerGraph} that these nodes are on the periphery of the network with lower connectivity compared to the rest of the graph. 

We further examine the cluster of reviewers in NIPS 2015 which is not well connected with the rest. In Figure~\ref{FigureClusterSubjects}, we plot the decomposition of this set in terms of the primary subject areas indicated by the reviewers. Our analysis reveals that a bulk of this cluster comprises a single subject area---reinforcement learning. Conversely, 50 out of 78 reviewers who identified their primary subject area as reinforcement learning lie in this cluster. 
All in all, graph connectivity issues of this form can lead to increased noise or bias in the overall decisions. Our main message for future conferences is to employ such methods of graph analysis in order to catch issues of this form \emph{at a global level} (not just local to individual ACs) before the reviews are assigned.

\begin{Summary}

\keyobservations{\item A cluster of papers and reviewers primarily in the reinforcement learning area are not well connected to the remaining papers and reviewers in the NIPS 2015 reviewer assignments. We did not find any such separated cluster in NIPS 2016.}

\actionitems{\item Use graph-theoretic techniques to check global structure of graph for reviewer assignment.}

\openquestions{\item Design principled graph-theoretic techniques, tailored specifically to the nuances of peer-review graphs, to verify soundness of reviewer assignments.}
\end{Summary}

\newsubsection{Review-score distribution and mismatches in calibration}
\label{SecScoreDistribution}

Recall from Section~\ref{SecReviewCriteria} that in the review process, for each feature, the reviewers were asked to provide a score on a scale of 1 to 5. Specifically, they were asked to provide a score of 5 for submissions they considered as being in the top $0.1\%$, a score of 4 for submissions that they deemed to be in the top $3\%$, and a score of 3 for submissions they deemed to be in the top $30\%$. In this section, we compare the actual empirical distribution of reviewer scores with the distribution prescribed in the guidelines to reviewers. 

\begin{figure}[t!]
\setlength{\belowcaptionskip}{-10pt}
\centering
\hspace{-.43cm}\begin{tabular}{c c c}
\incl{0.48}{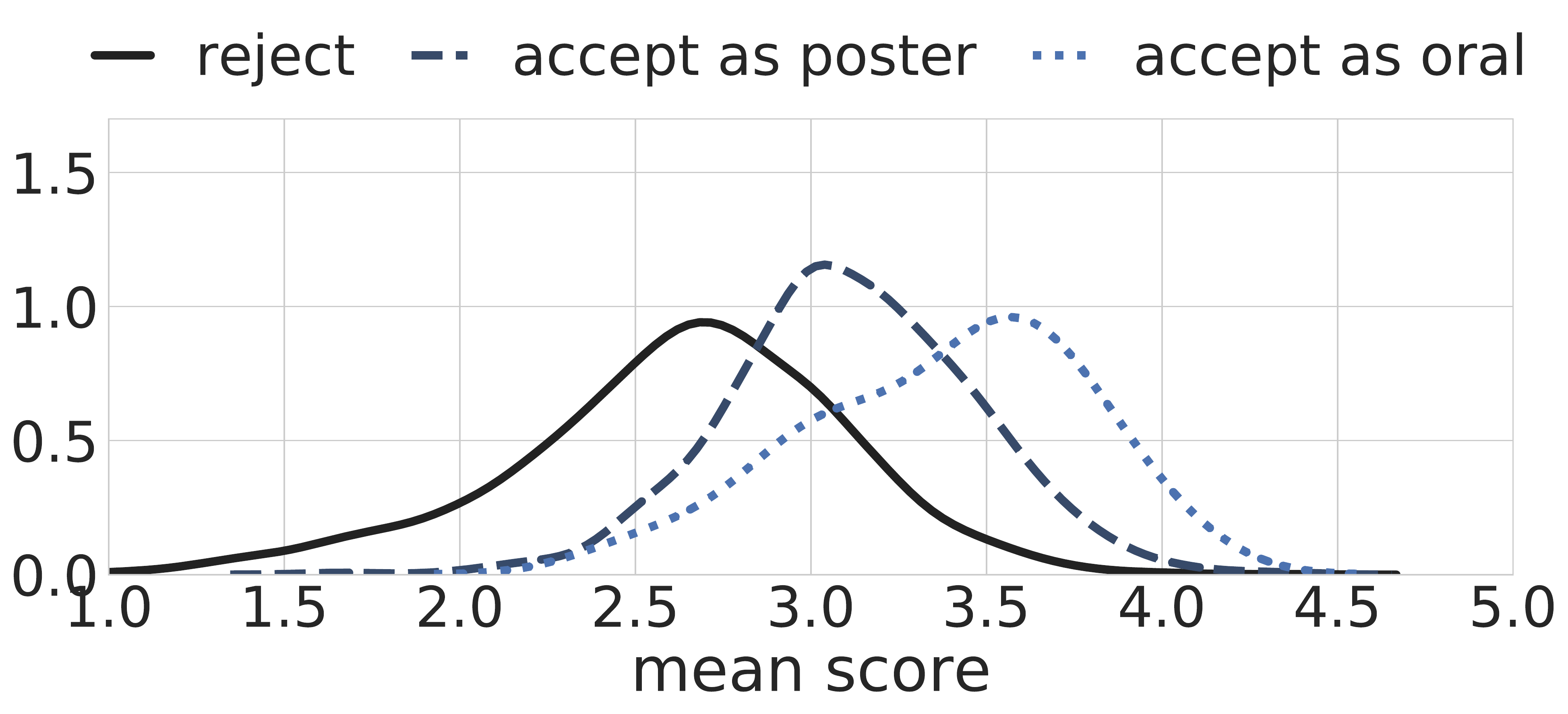} & 
&
\hspace{-.3cm}\incl{0.48}{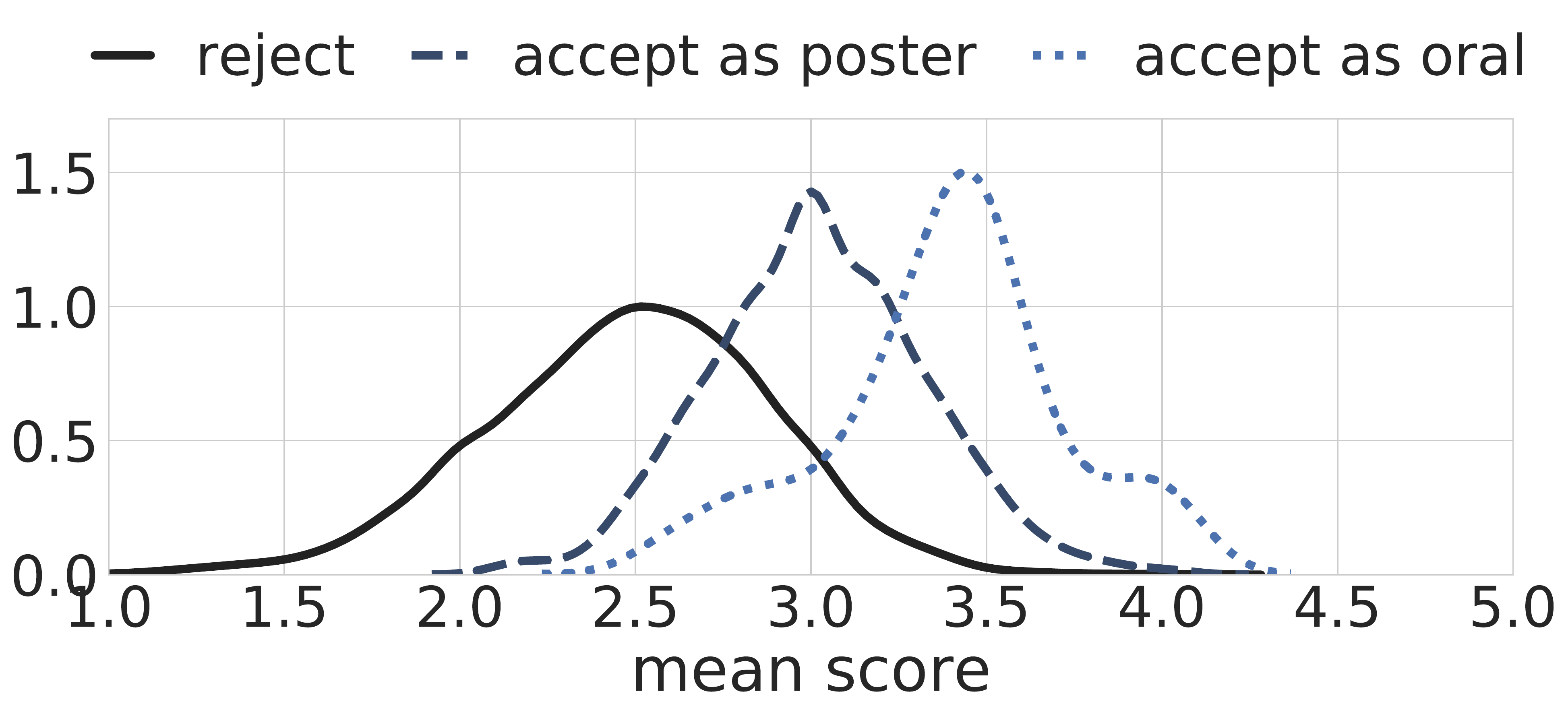} \\
(a) clarity &  & \hspace{-.3cm}(b) impact\\
\incl{0.48}{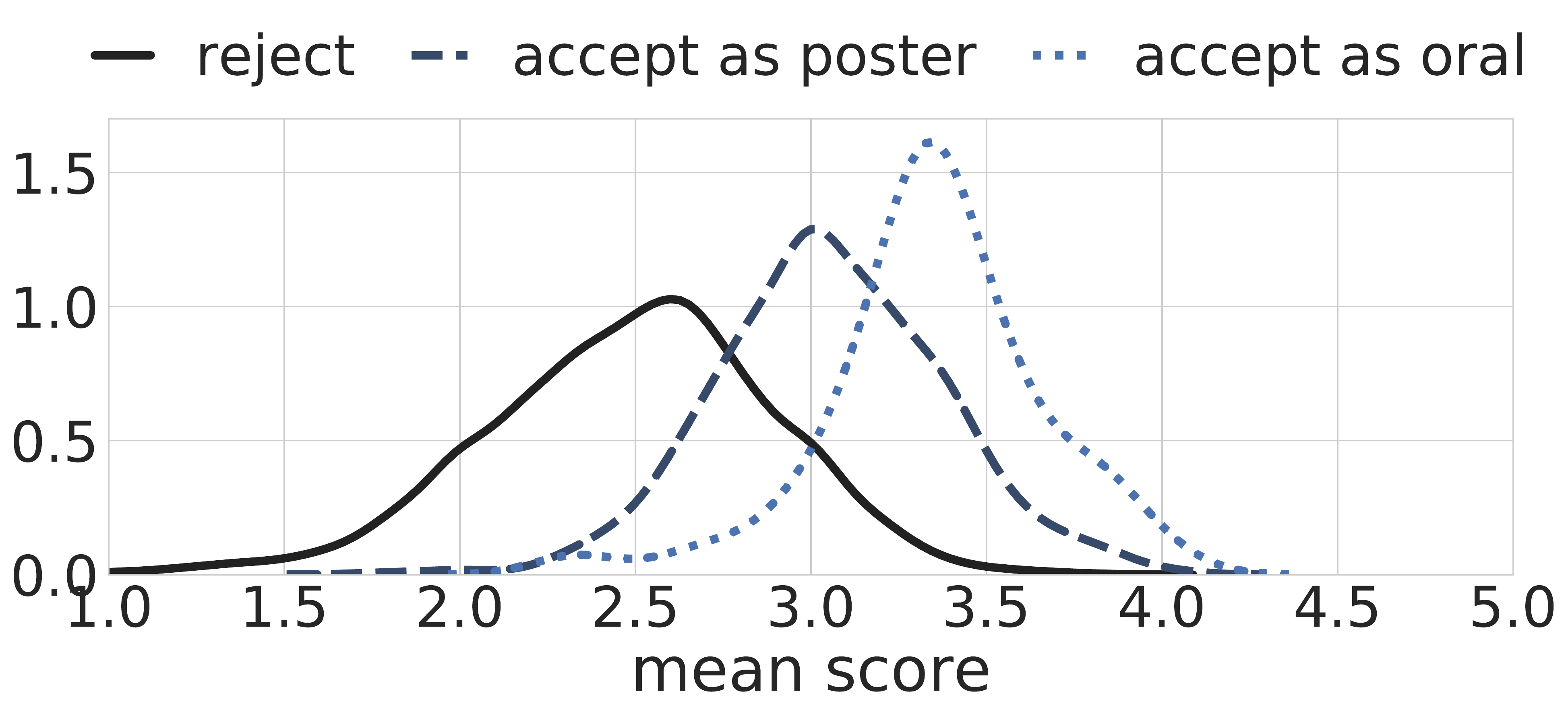} &
 &
\hspace{-.3cm}\incl{0.48}{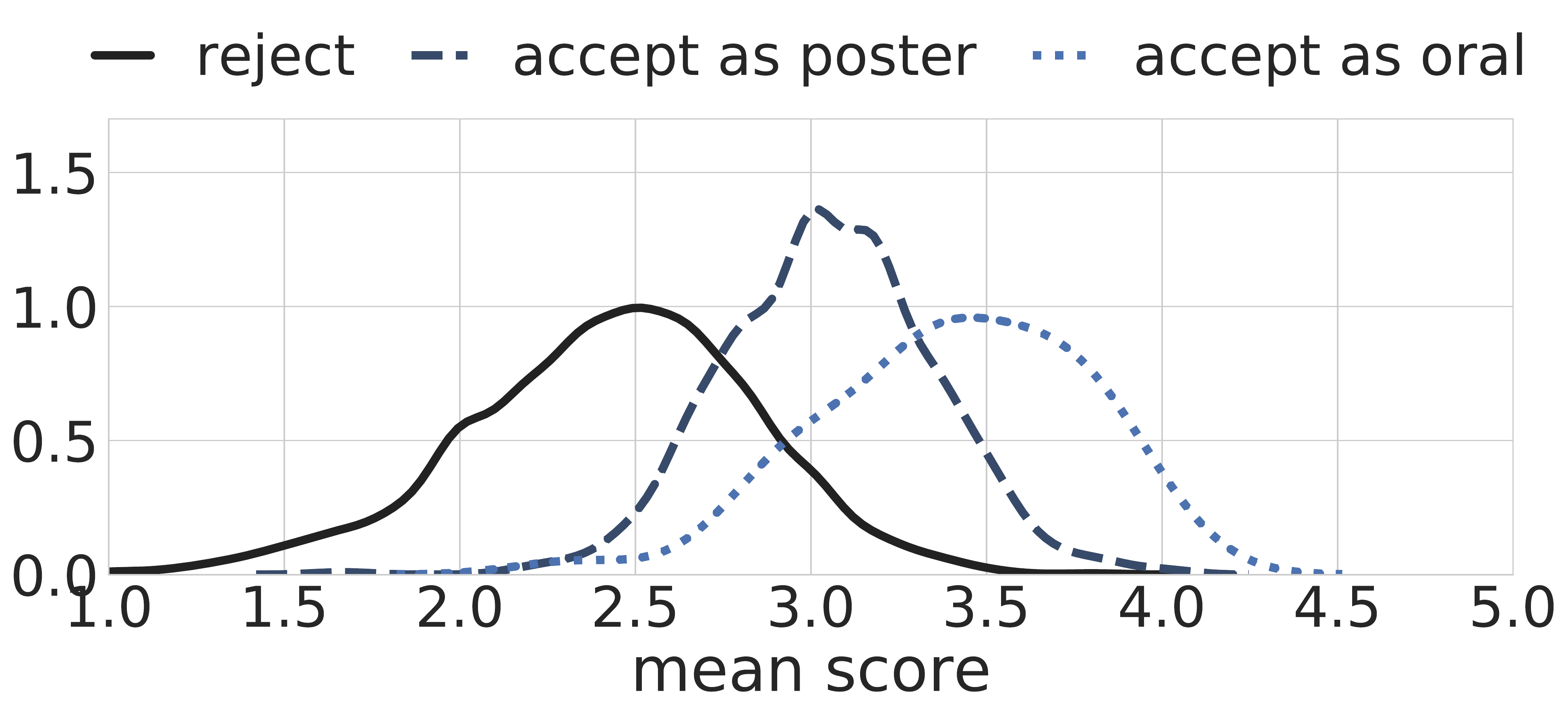} \\
(c) novelty &  & \hspace{-.3cm}(d) quality\\
\end{tabular}
\caption{Distribution of the mean value (across reviewers) of the score per paper for different features, separated according to the final decisions. 
\label{FigScoreDistribution}}
\end{figure}

We begin by computing the distribution of the mean value (across reviewers) of the score per paper for different features,  separated according to the final decisions. We plot these distributions in Figure~\ref{FigScoreDistribution} for each of the four features of clarity, impact, novelty, and quality separately.

At first glance, these histograms and numbers look quite reasonable. However, what was surprising to us was the percentage of papers that received any particular score -- see Table~\ref{TabCalibrationMismatch}. Even though the reviewers were asked to give a paper a score of 3 (poster level) or higher only if they think the paper lies in the top $30 \%$ of all papers, nearly $60\%$ of the scores were 3 or higher. Similar effects occurred for scores 4 and 5.

\begin{table}[b!]
\centering
\begin{tabular}{|c|c|c|c|c|c|}
\hline
 & \begin{tabular}{@{}c@{}}1 \\ (low or very low)\end{tabular}  & \begin{tabular}{@{}c@{}}2 \\ (sub-standard)\end{tabular}  & \begin{tabular}{@{}c@{}}3 \\ (poster level: \\ top $30\%$)\end{tabular}  & \begin{tabular}{@{}c@{}}4 \\ (oral level: \\ top $3\%$)\end{tabular}   & \begin{tabular}{@{}c@{}}5 \\ (award level: \\top $0.1\%$)\end{tabular} \\
 \hline
 Impact & 6.6\,\% & 36.4\,\% & 45.9\,\% & 10.7\,\% & 0.4\,\% \\
 Quality & 6.7\,\% & 38.3\,\% & 45.0\,\% & 9.6\,\% & 0.4\,\% \\
 Novelty & 6.4\,\% & 35.0\,\% & 48.4\,\% & 9.8\,\% & 0.4\,\% \\
 Clarity & 7.1\,\% & 28.1\,\% & 48.9\,\% & 14.7\,\% & 1.2\,\% \\
 \hline
\end{tabular}~\\~\\
\caption{Distribution of the reviews according to the provided scores for each of the four features. The column headings indicate the guidelines that were provided to the reviewers. Observe that the percentage of reviews providing scores of $3$, $4$ or $5$ is considerably higher than the requested values.\label{TabCalibrationMismatch}}
\end{table}

One possible explanation for this phenomenon is that there were a large number of high-quality submissions to NIPS 2016. Such an improvement in quality has obvious upsides such as uplifting the overall experience of the conference. The downside is that the burden on selecting the accepted papers among all those good submissions is with the area chairs, who now still had to reduce the $60\%$ good papers to $23\%$ accepted papers. A second possible explanation is that the reviewers were not calibrated that well with respect to the paper quality. In either case, we understand that this obviously led to the frustration of many authors, whose papers received good scores but were rejected. 

In addition to scores for the four features, the reviewer could also indicate whether the paper had a ``fatal flaw''. We observe that \numberchecked{$32\%$} of all papers were flagged to have a ``fatal flaw'' by at least one reviewer. 

\begin{Summary}
\keyobservations{\item The fraction of reviews with high ratings is significantly higher than what was asked from the reviewers. For instance, nearly $60\%$ of scores are 3 or higher even though reviewers were asked of scores of 3 or higher only when they thought the paper was in the top $30\%$ of submissions.
}

\actionitems{ 
\item If eliciting ratings, do not use numbered scales (that is, do not use ``1'', ``2'', …). Alternatively, one may employ other means of elicitation such as rankings.

\item When making reviews visible to authors, show the percentile with respect to the data instead of absolute scores,  e.g., provide feedback of the form ``your paper is in the top 40\% of all submitted papers in terms of novelty...''

\item Include an expert in elicitation, survey methodology or user interface design to help to design what and how to ask~\citep{ohagan2006uncertain}.
}

\openquestions{\item Since each reviewer reviews only a small subset of the submitted papers, how to calibrate the reviews?
\item What is the best interface for eliciting reviewer responses?
\item What is the best way to present the review results to authors in order to provide most useful feedback and minimizing distress?
}
\end{Summary}

\newsubsection{Behavior of different pools of reviewers}
\label{SecDifferentReviewers}

In this section, we compare the reviews provided by the volunteer (pool 2) reviewers to those provided by the invited (pool 1) reviewers. The inclusion of volunteer reviewers has two important benefits: (a) It increases the transparency of the review process. (b) Volunteer reviewers may be new today but in 2 years down the line they will gain experience and become useful to accommodate the massive growth of the conference. Given these benefits of including volunteer reviewers, this analysis looks for any systematic differences between the review scores provided by the two pools of reviewers.

\begin{figure}[!b]
\centering
\subfloat[invited and volunteer reviewers]{\includegraphics[height = 1.3in]{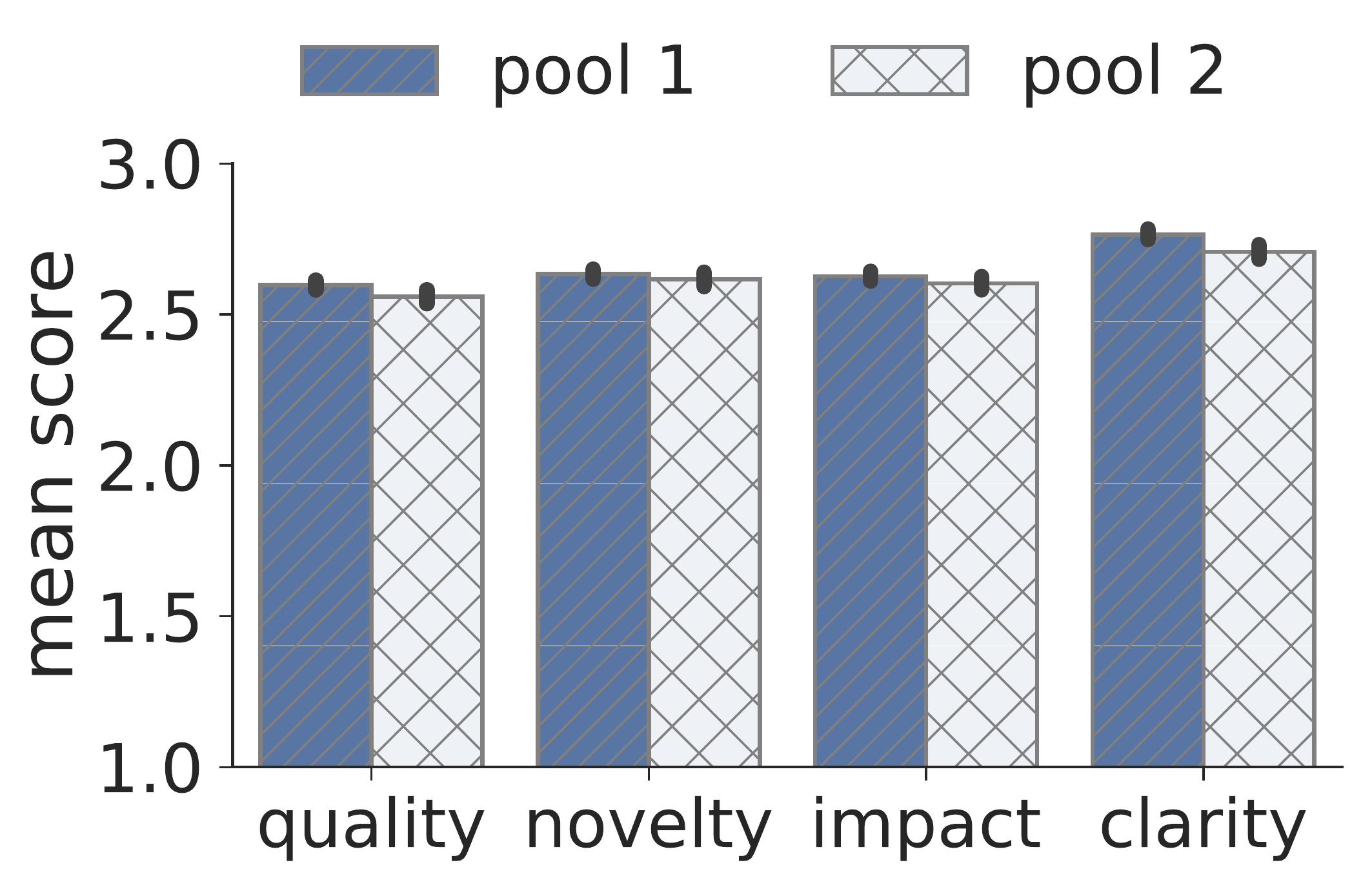}}\qquad
\subfloat[senior, junior, and student reviewers]{\includegraphics[height = 1.3in]{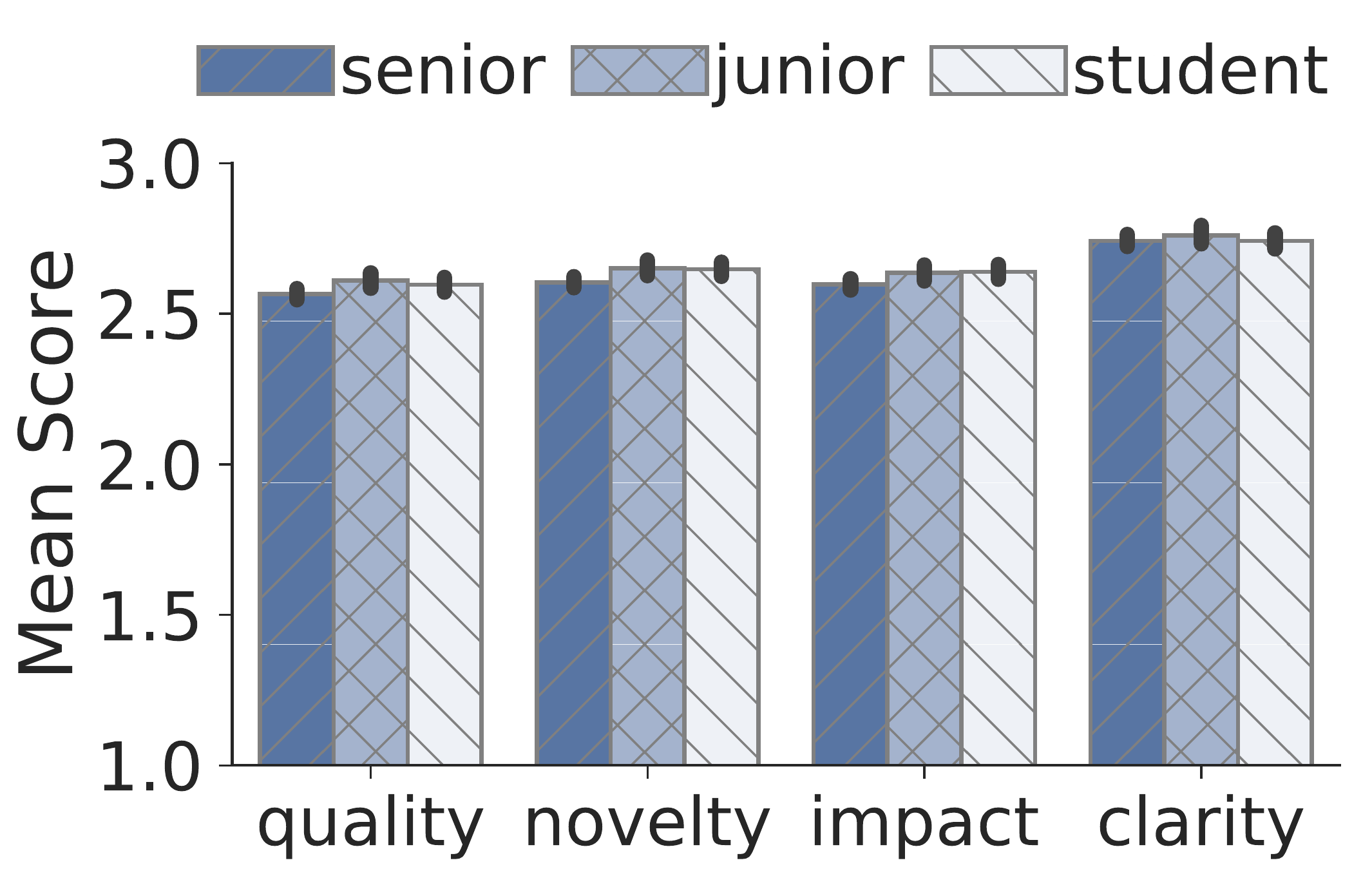}}
\caption{Mean of scores provided for different features grouped by different reviewer types. 
\label{FigScorePoolType}}
\end{figure}

\xhdr{Mean scores} 
Junior reviewers are often perceived to be more critical than senior reviewers~\citep{apa2007gradstudentreviewers, toor2009readinggradstudent}. As~\citet{apa2007gradstudentreviewers} notes, \emph{``You submit your manuscript and then just pray it doesn't get sent to a junior faculty member -- young faculty are merciless!''} In this section, we examine this hypothesis in the NIPS 2016 reviews. In Figure~\ref{FigScorePoolType}, we plot the mean score provided by each group of reviewers for each individual feature. We apply a t-test on observed scores and compute the effect size to examine if there is a statistically significant difference in the underlying means of the scores provided by different categories of reviewers. For Pool 1 vs Pool 2, this analysis shows only clarity to have a statistically significant difference between the two pools after accounting for multiple testing. Specifically, the p-values (before accounting for multiple testing) and effect sizes for the four features are: \numberchecked{novelty p=$0.2143$, \effectsize{$0.0264$}, quality p=$0.0061$, \effectsize{$0.0581$}, impact p=$0.0961$, \effectsize{$0.0353$}, and clarity p=$1.91 \times 10^{-04}$, \effectsize{$0.0788$}}. Sample sizes for Pool 1 and Pool 2 reviews are \numberchecked{$9244$ and $4430$} respectively.  

A similar analysis between senior researchers (e.g., faculty), junior researchers (e.g., postdocs), and PhD students reveals no significant  difference between these categories. Specifically, the p-values (before accounting for multiple testing) and effect sizes for senior researcher vs. junior researchers for the four features are: \numberchecked{quality p=$0.0071$, \effectsize{$-0.0662$}, novelty p=$0.0037$, \effectsize{$-0.0704$}, impact p=$0.0199$, \effectsize{$-0.0569$}, and clarity p=$0.3064$, \effectsize{$-0.0253$}};
\numberchecked{for junior researcher vs. students: quality p=$0.4662$, \effectsize{$0.0164$}, novelty p=$0.8247$, \effectsize{$0.0049$}, impact p=$0.8733$,\effectsize{$-0.0036$}, and clarity p=$0.3529$, \effectsize{$0.0209$}};
for senior researcher vs. students: \numberchecked{quality p=$0.0440$, \effectsize{$-0.0454$}, novelty p=$0.0499$, \effectsize{$-0.0629$}, impact p=$0.0076$, \effectsize{$-0.0601$} and clarity p=$0.9968$, \effectsize{$0.00009$}}.  The sample sizes for senior, junior and student reviews are: \numberchecked{$6335$, $3938$, and $3354$} respectively. This analysis excludes $47$ reviews by reviewers who did not identify themselves as any of the above categories.

\begin{figure}[!b]
\centering
\subfloat[invited and volunteer reviewers\label{fig:conf_pool}]{\qquad\includegraphics[height = 1in]{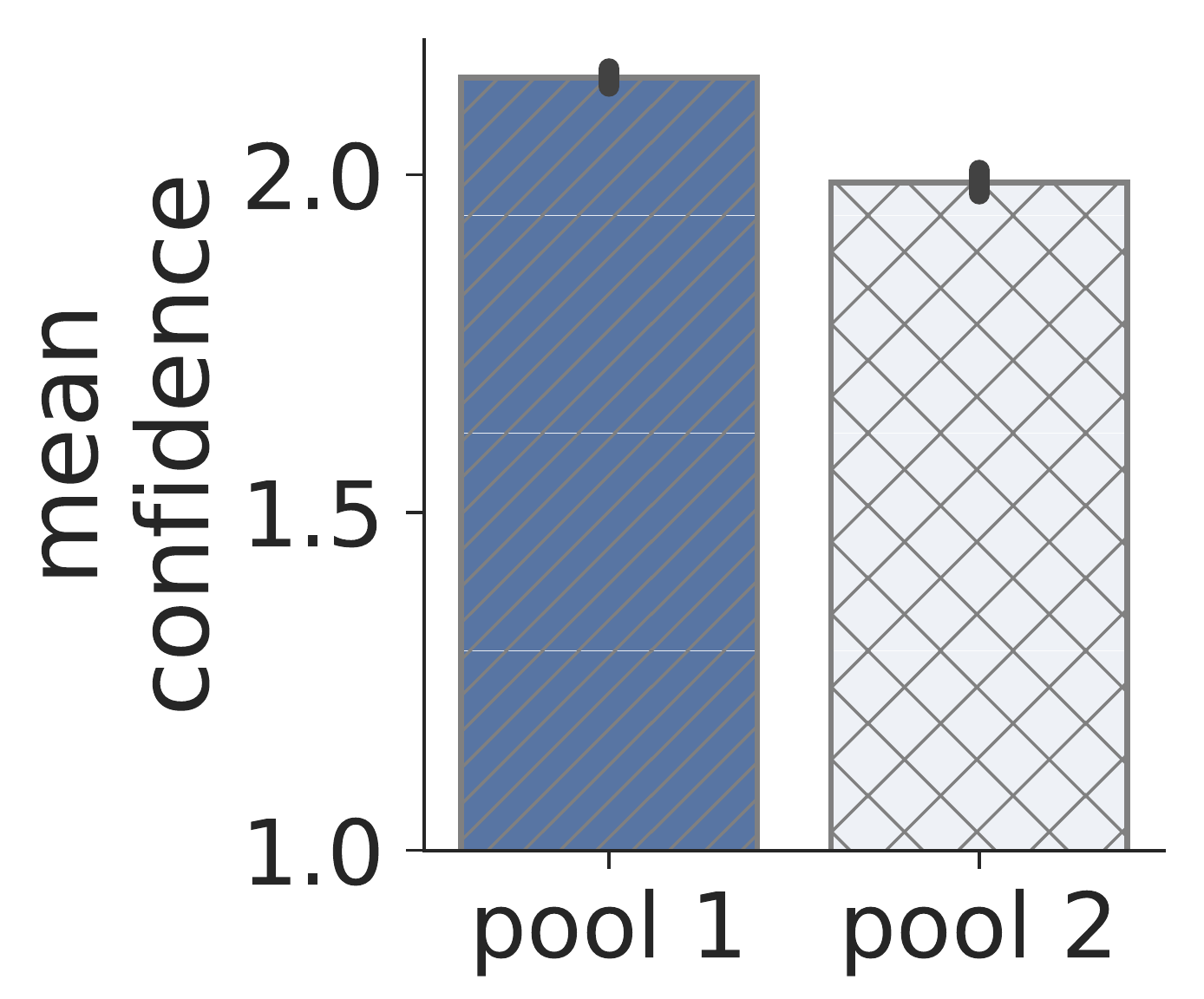}\qquad}\qquad
		\subfloat[senior, junior, and student reviewers\label{fig:conf_st}]{\qquad\includegraphics[height = 1in]{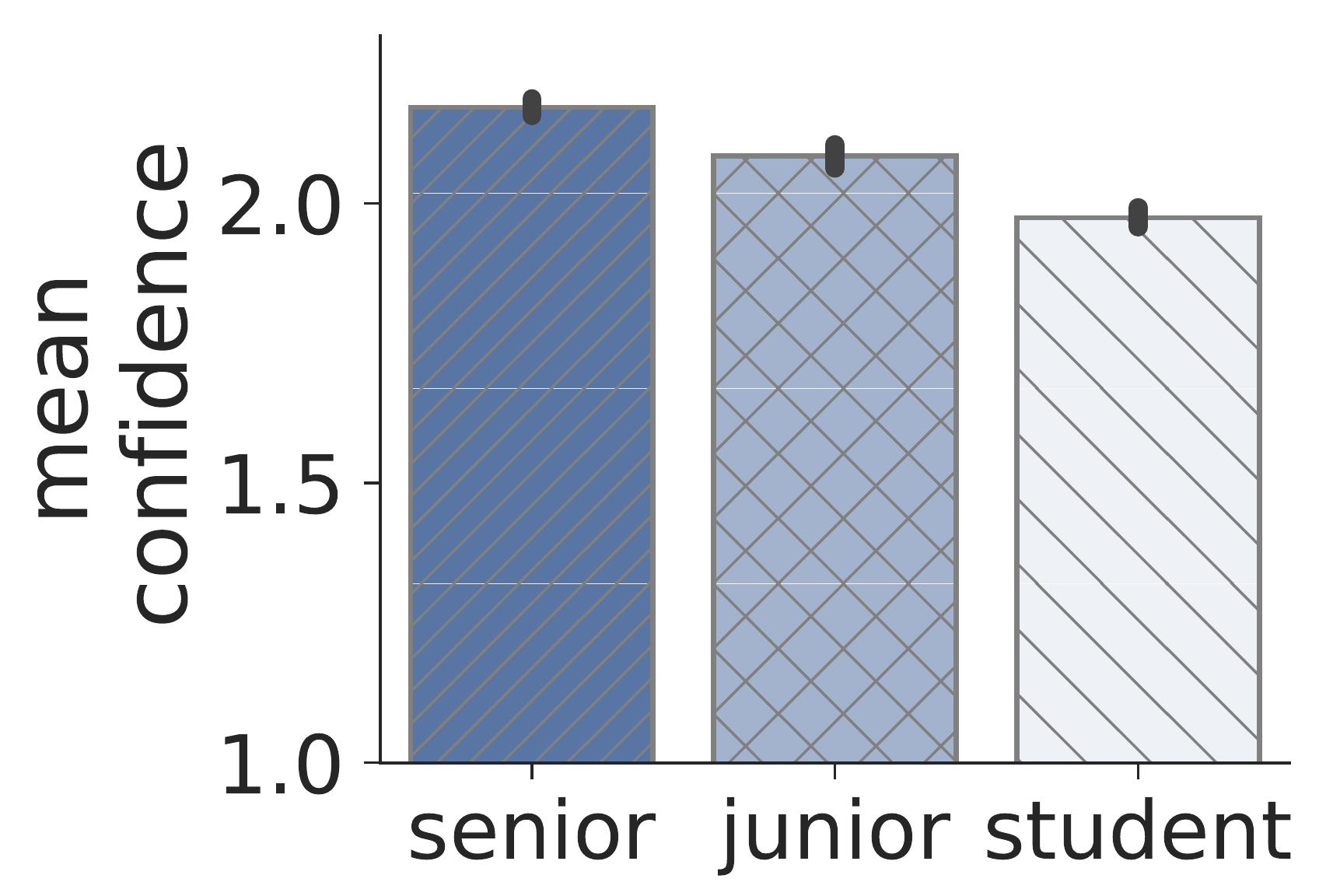}\qquad}
	\caption{Self-reported confidence grouped by different reviewer types. 
	\label{FigConfidencePoolType}}
\end{figure}

 \begin{figure}[!b]
 \ifarxiv{\!\!\!\!\!\!}\setlength{\tabcolsep}{0.01em}
 	\centering
 	\subfloat[novelty]{\includegraphics[height = 0.99in]{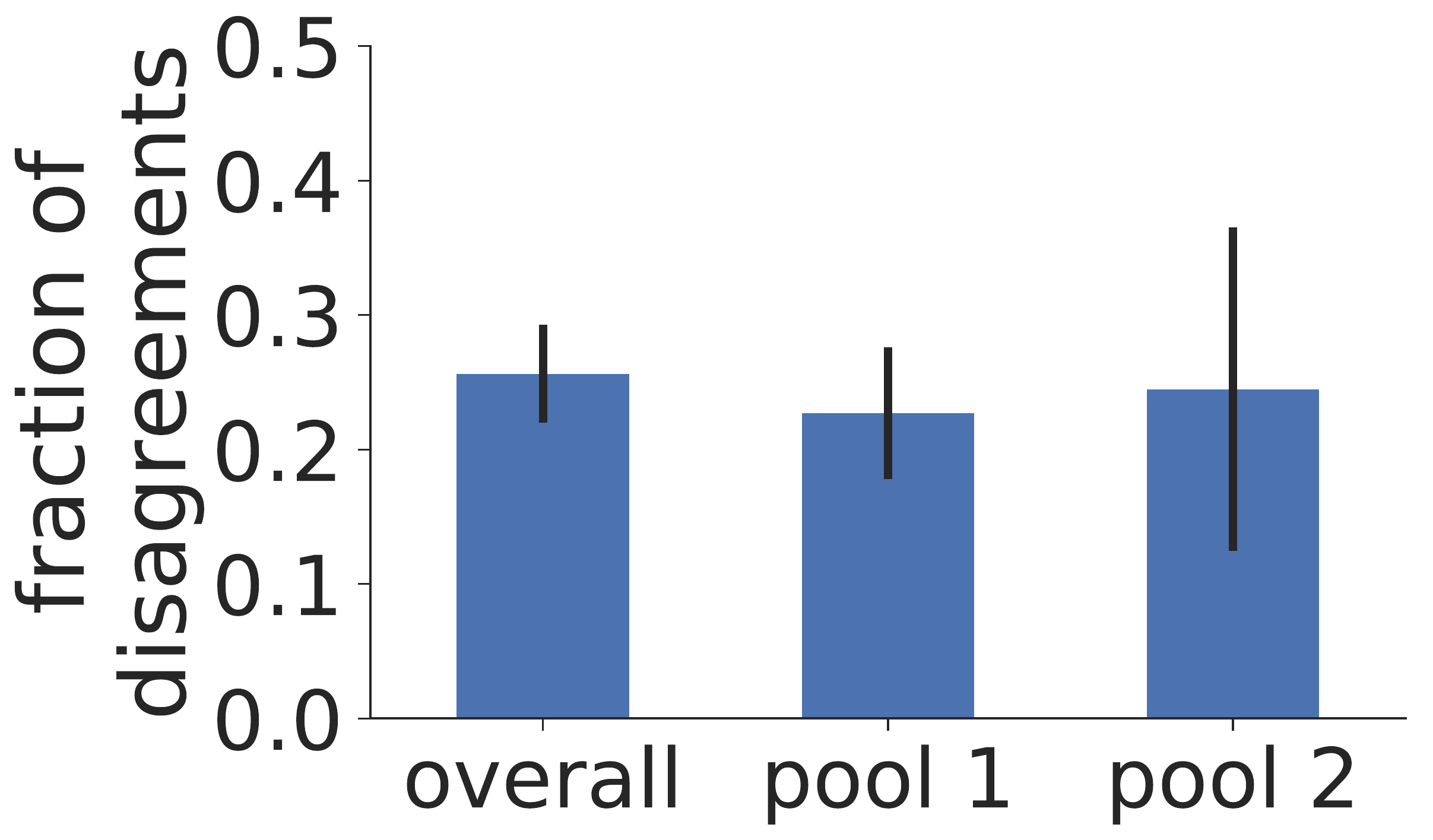}}
 	\subfloat[quality]{\includegraphics[height = 0.99in]{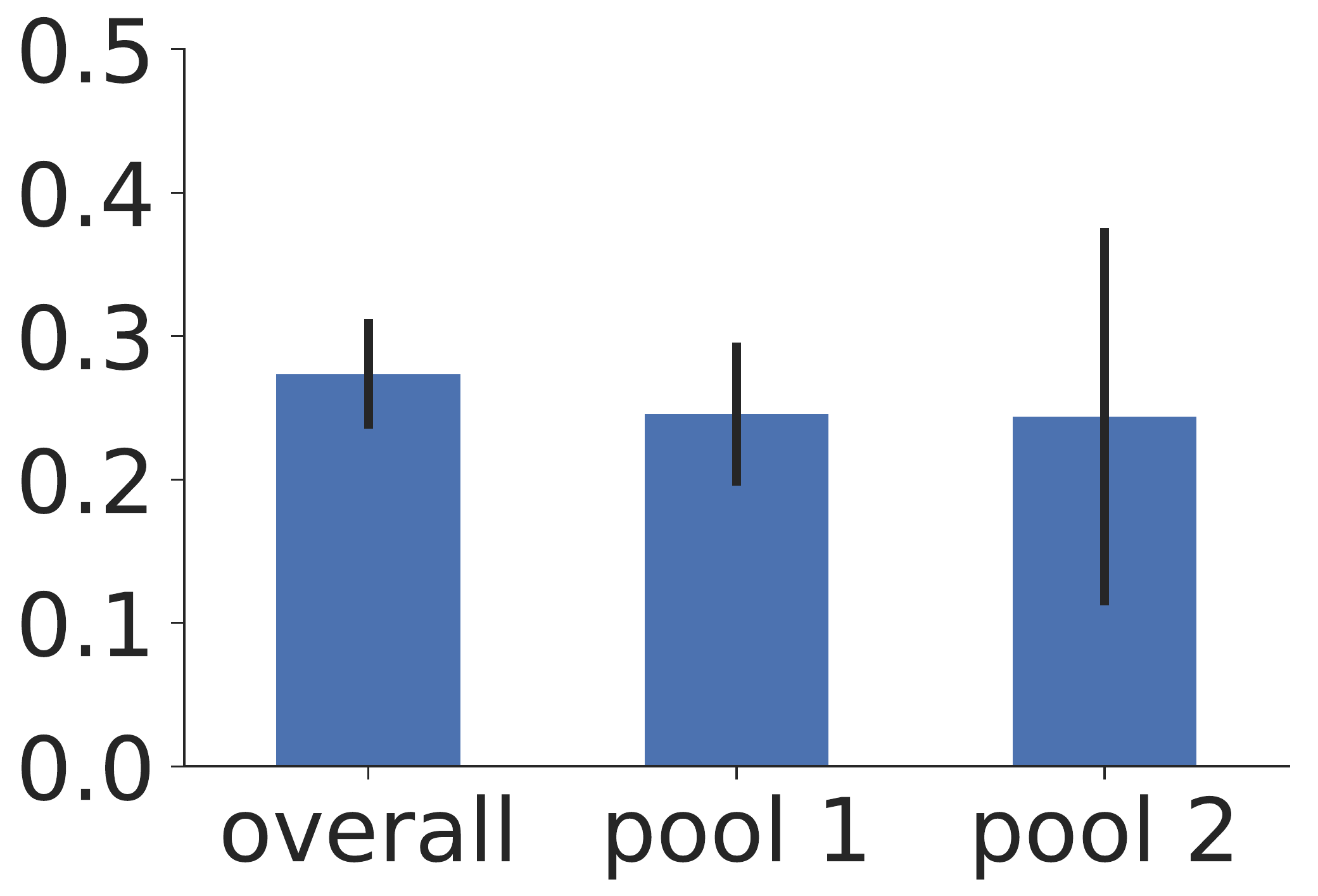}}
 	\subfloat[impact]{\includegraphics[height = 0.99in]{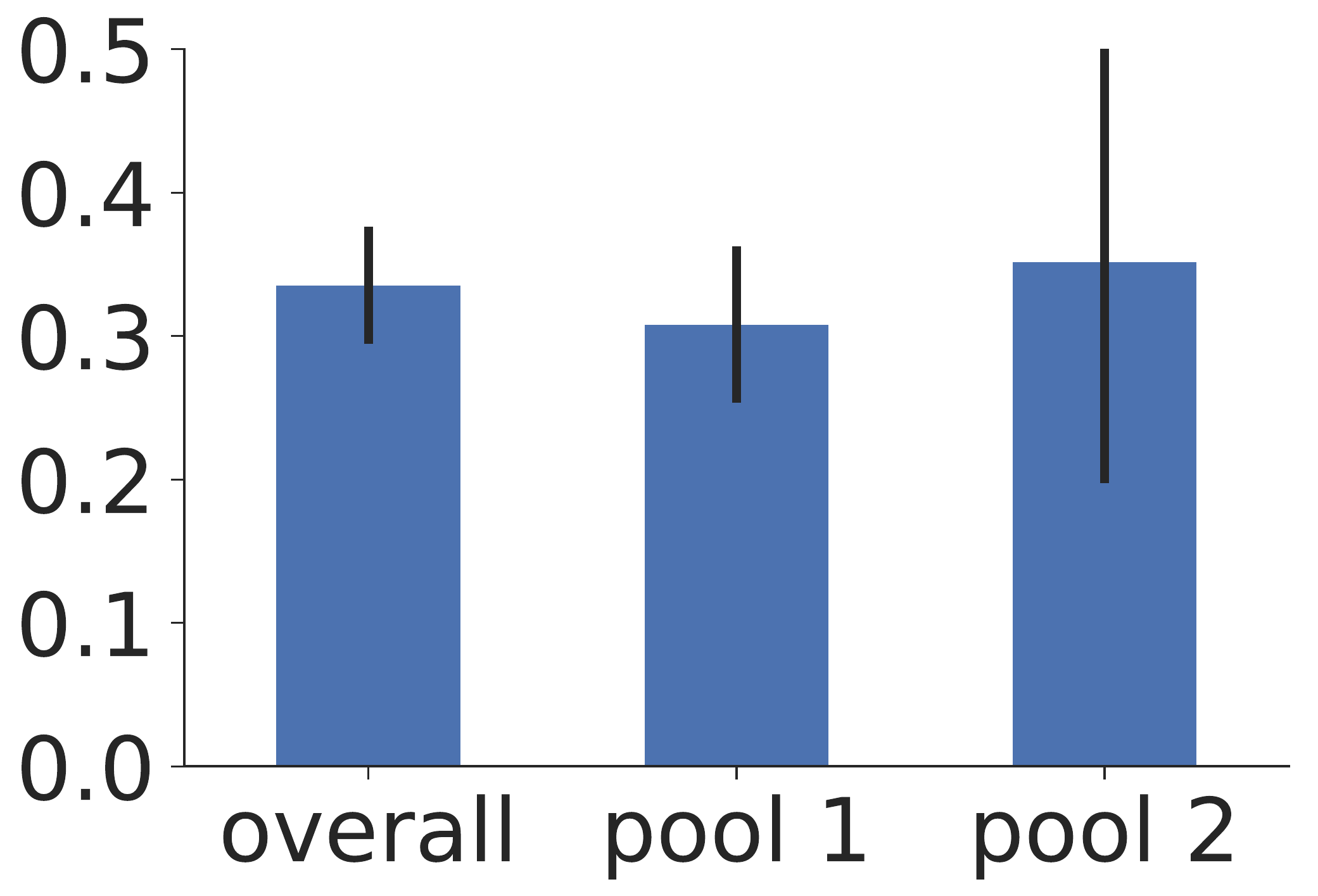}}
 	\subfloat[clarity]{\includegraphics[height = 0.99in]{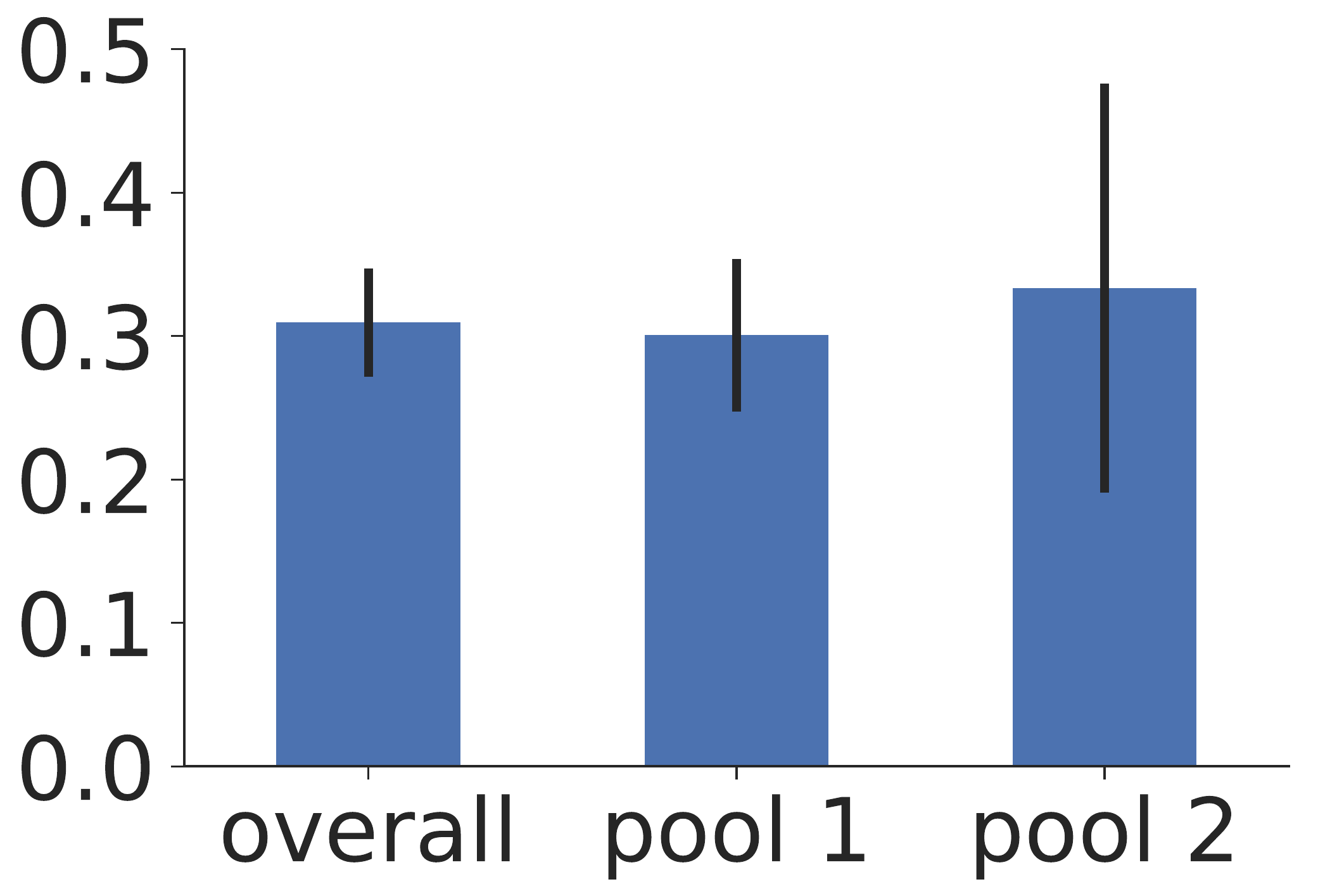}}
 	\caption{Proportions of inter-reviewer disagreements on each feature.
 	\label{FigureCardInterWorkerAgreeAfter}
 	}
 \end{figure}

\xhdr{Self-reported confidence}
We next study the difference in the self-reported confidence among different groups of reviewers. The mean value of reported confidence is plotted in Figure~\ref{FigConfidencePoolType}. In this case, we see a statistically significant correlation between seniority and self-reported confidence. Following are p-values (before accounting for multiple testing) and corresponding effect sizes: senior vs. junior researcher: \numberchecked{p=$4.1683 \times 10^{-11}$, \effectsize{$0.1604$}, senior researcher vs. PhD student: p=$3.308 \times 10^{-57}$, \effectsize{$0.3577$} and junior researcher vs. PhD student: p=$8.074 \times 10^{-15}$, \effectsize{$0.1758$}}. We observe a similar difference in confidence score and effect size between pool 1 and pool 2 reviewers: \numberchecked{p=$3.9679 \times 10^{-44}$, \effectsize{$0.2943$}}.

\xhdr{Consistency}
We now study the consistency within reviewers of pool 1 (invited), and within reviewers of pool 2 (volunteer). The consistency captures the amount of variance or disagreements in the reviews provided by that pool. As noted by~\citet{ragone2013peer}, \emph{``the disagreement among reviewers is a useful metric to check and monitor during
the review process. Having a high disagreement means, in some way, that the
judgment of the involved peers is not sufficient to state the value of the contribution
itself. This metric can be useful to improve the quality of the review process...''}

Concretely, consider any pair of reviewers within a given pool, any pair of papers that is reviewed by both the reviewers, and any feature. We say that this pair of reviewers agrees on this pair of papers (for this feature) if both reviewers rate the same paper higher than the other; we say that this pair disagrees if the paper rated higher by one reviewer is rated lower by the other. Ties are discarded. We count the total number of such agreements and disagreements within each of the two pools.

Figure~\ref{FigureCardInterWorkerAgreeAfter} plots the fraction of disagreements within each of the two pools for the cardinal scores. At this aggregate level, we do not see enough difference to conclusively rate any one pool's intra-pool agreement above the other, and this conclusion is also confirmed by an absence of a statistically significant difference in the proportion of agreements within pool 1 from the proportion of agreements within pool 2. Specifically, for the Pearson's chi-squared test and effect sizes of pool 1 vs. pool 2, the results for  the four features (before accounting for multiple testing) are:\numberchecked{ novelty p=$0.9269$ \effectsize{ -0.0426}, quality p=$0.8648$, \effectsize{0.0039}, impact p=$0.7296$, \effectsize{-0.0936}, and clarity p=$0.8029$, \effectsize{-0.0709}}.  The total sample sizes for the three categories of overall, pool 1 and pool 2 respectively across the four features are: novelty \numberchecked{554, 282 and 49; quality 523, 285 and 41; impact 513, 276 and 37; and  clarity 572, 286 and 42}. Section~\ref{SecOrdinal} presents similar consistency results for the two pools in the ordinal rankings. (We also attempted to run this analysis restricted to the \toptwok papers, but this restriction results in a very low sample complexity and hence underpowered tests.)

\xhdr{Participation in discussions} One fact that caught our attention was the amount of participation in the discussion by the different reviewer groups: senior reviewers take much more active roles in the discussions than junior researchers. Please see Section~\ref{SecDiscussionParticipation} for details, where we provide a more detailed study of the discussion phase.

\begin{Summary}
\keyobservations{\item We find no evidence of a critical bias of junior reviewers (except for the ``clarity'' feature).
\item Self-reported confidence correlates with seniority.
\item Volunteer reviewers yield benefits of scalability and transparency, with no observable biases and a similar inter-reviewer agreement as the invited pool. These reviewers can soon be an asset in dealing with the rapid growth of conferences such as NIPS. 
}

\actionitems{\item Continue to include volunteer reviewers.
}

\openquestions{\item How do we make most effective use of volunteer reviewers in a manner that authors can trust, reduces randomness in the peer-review process, and trains junior reviewers effectively?
}
\end{Summary}

\newsubsection{Rebuttals and discussions} 
\label{SecDiscRebuttals}

This section is devoted to the analysis of the rebuttal stage and the participation of reviewers in discussions. 
We begin with some summary statistics. The authors of 2188 papers submitted a rebuttal.

There were a total of \numberchecked{12154} reviews that came in before the rebuttals started, and with some more reviews received  after the rebuttal round, the total number of \tuple{reviewer, paper} pairs eventually ended up being \numberchecked{13674}. Among the \numberchecked{12154} reviews that were submitted before the rebuttal, the scores of only \numberchecked{1193} of them changed subsequent to the rebuttal round. These changed review scores were distributed among \numberchecked{886} papers. 

There were 842 papers for which no reviewer participated in the discussions, 339 papers for which exactly one reviewer participated, and 436, 376, 218, 135 and 49 papers for which 2, 3, 4, 5 and 6 reviewers participated respectively. There were a total of 5255 discussion posts, and 4180 of the 13674 \tuple{reviewer, paper} pairs participated in the discussions.


\subsubsection{Who participates in discussions?}
\label{SecDiscussionParticipation}

We compare the amount of participation of various groups of reviewers in the discussion phase of the review process. 

\begin{figure}[b!]
\centering
\subfloat[invited and volunteer reviewers]{
\incl{0.39}{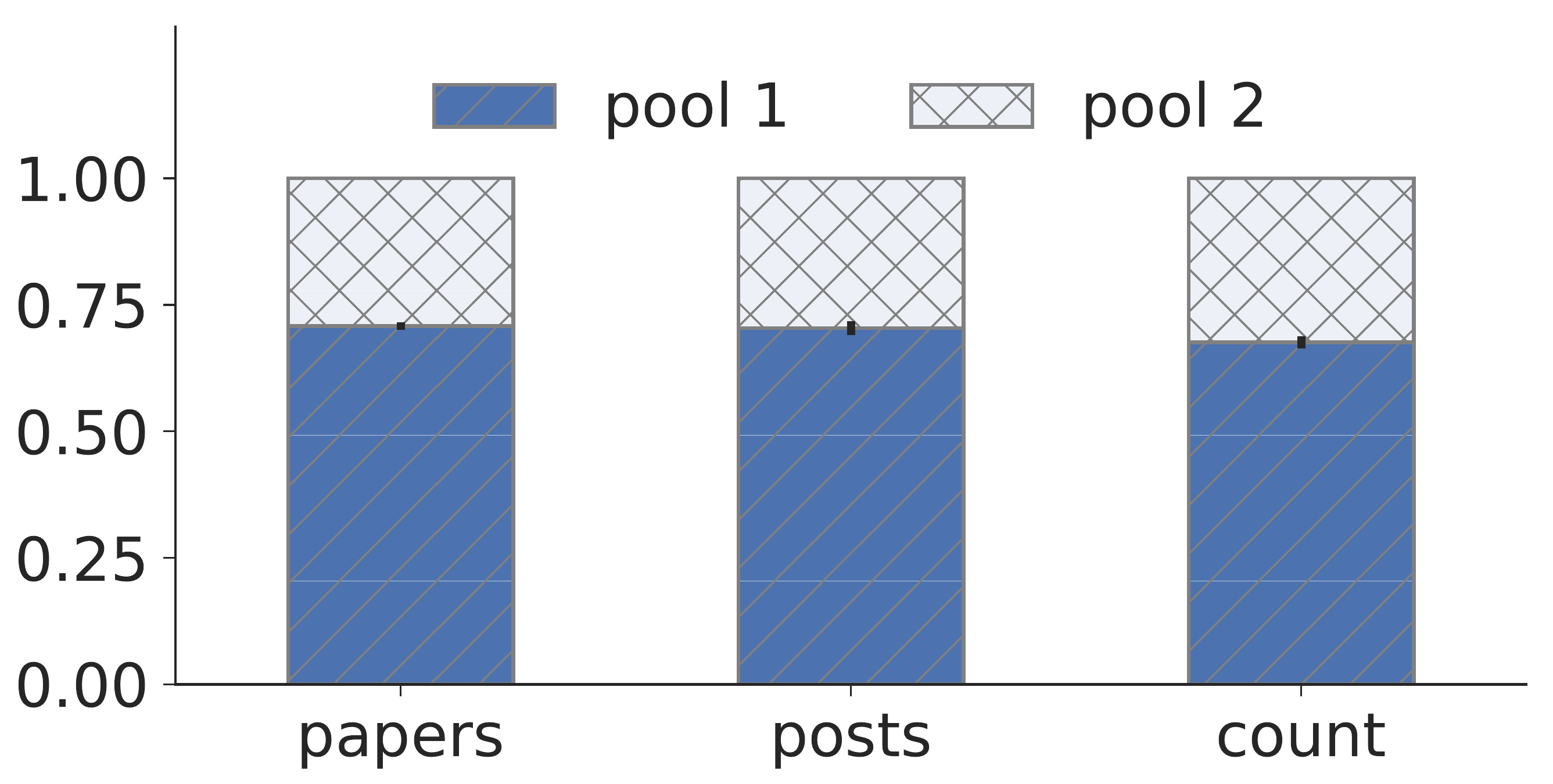}}\qquad
\subfloat[student and non-student reviewers]{\incl{0.39}{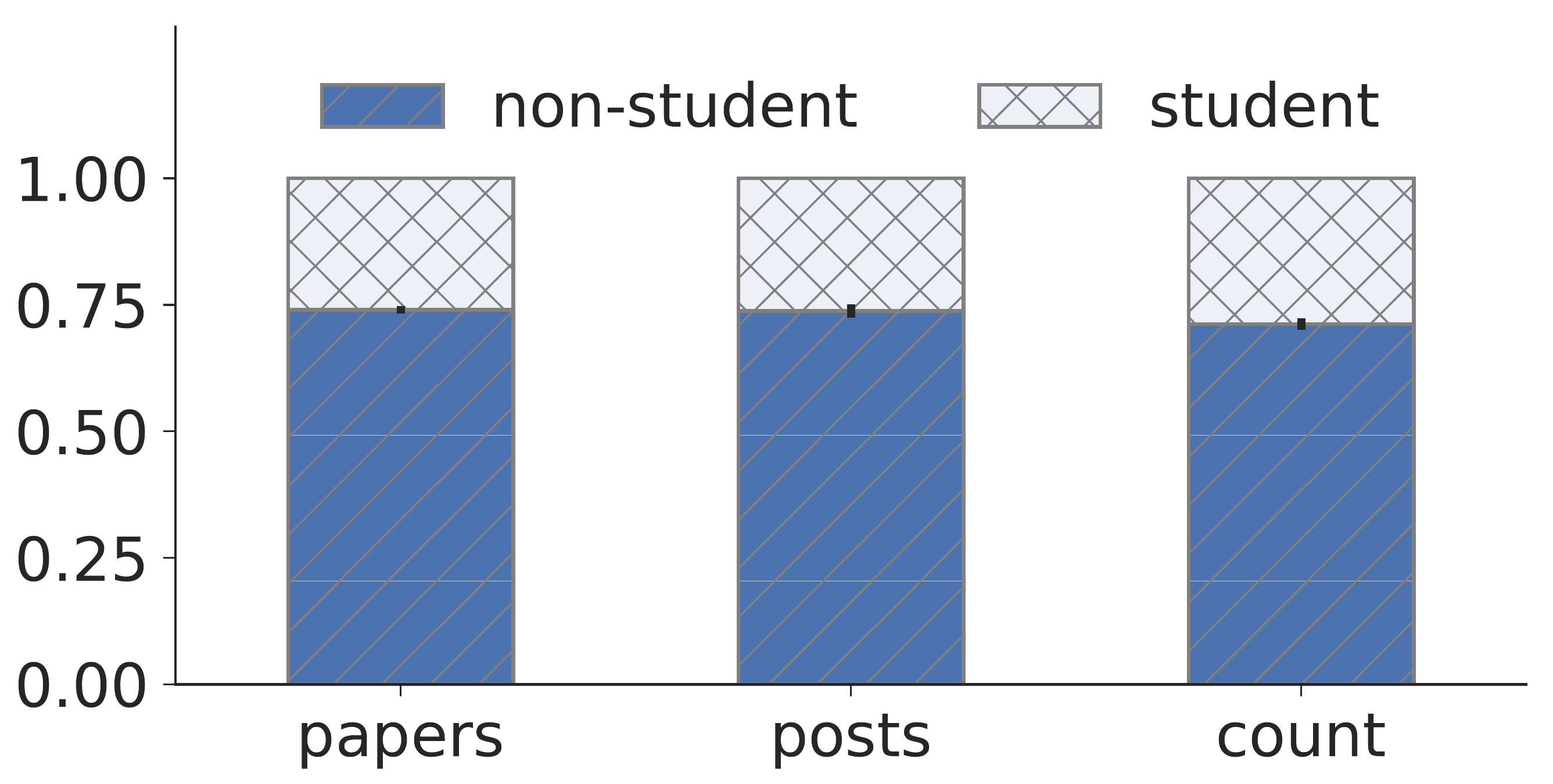}}
\caption{Proportions of contributions from different types of reviewers in discussions (``posts'' and ``papers'') and the total number of such reviewers (``count'').}
\label{fig:discussion}
\end{figure}

\xhdr{Pool 1 (invited) versus pool 2 (volunteer) reviewers}
We compare the participation of the reviewers in two pools in the discussions as follows, and plot the results in Figure~\ref{fig:discussion}(a). In order to set a baseline, we first compute the total number of \tuple{pool 1 reviewer, paper} pairs and the total number of \tuple{pool 2 reviewer, paper} pairs -- these counts are computed irrespective of whether the reviewer participated in the discussions or not. We plot the proportions of these counts as the ``count'' bar in the figure. Next we compute the total number of posts that were made by pool 1 reviewers and the total number of posts that were made by pool 2 reviewers -- the resulting proportions are plotted as the ``posts'' bar in the figure. Finally, we compute the total number of \tuple{pool 1 reviewer, paper} pairs in which that reviewer put at least one post in the discussion for that paper, and the total number of \tuple{pool 2 reviewer, paper} pairs in which that reviewer put at least one post in the discussion for that paper. We plot the two proportions in the ``papers'' bar.  The total sample sizes for the  categories of counts, posts and papers are \numberchecked{13674, 5255 and 4180} respectively.

We tested whether the mean number of posts per \tuple{reviewer, paper} pair is identical for the two pools of reviewers. For the null hypothesis that the means are identical for the two pools of reviewers, the t-test yielded a p-value of \numberchecked{p = $1.36 \times 10^{-4}$}. We also conducted this analysis for the restriction of papers to the \toptwok, and for this subset, the t-test yielded a p-value of \numberchecked{p = $9.458 \times 10^{-4}$}. We see a statistically significantly higher participation by the pool 1 reviewers as compared to the pool 2 reviewers in the discussions. However, the absolute amount of participation by either group is moderate at best, and
the effect sizes are small with \numberchecked{\effectsize{0.0704} and \effectsize{0.0894}} for analysis of all papers and \toptwok papers respectively.

\xhdr{Student versus non-student reviewers} 
We calculated the above three sets of quantities for student and non-student reviewers. Figure \ref{fig:discussion}(b) depicts the results. We tested whether the mean number of posts per \tuple{reviewer, paper} pair for the student reviewers is identical to the non-student reviewers. For the null hypothesis that the means are identical, the t-test yielded a p-value of \numberchecked{p = $3.016 \times 10^{-4}$}. We also conducted this analysis for the restriction of papers to the \toptwok, and for this subset, the t-test yielded a p-value of \numberchecked{p = $8.932 \times 10^{-4}$}. We see a statistically significantly higher participation by the non-student reviewers as compared to the student reviewers in the discussions. However, the total amount of participation by either group is not too large, and
the effect sizes are small with \numberchecked{\effectsize{0.0695} and \effectsize{0.0929}} respectively.


\begin{figure}[t!]
\centering
\includegraphics[width = \textwidth]{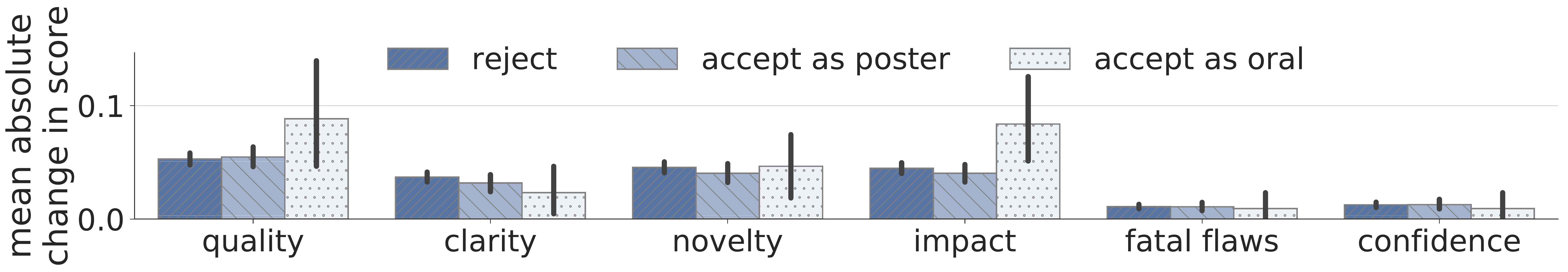}\\
\caption{Mean absolute value of the change in the scores from before the rebuttal round to the end of the discussion phase. \label{FigDiscussionAbsoluteScores}}
\end{figure}


\subsubsection{How do discussions change the scores?}
\label{SecRebuttalChange}

A total of \numberchecked{1193} out of \numberchecked{12154} reviews that were submitted before rebuttals changed subsequently. These changed reviews were distributed among \numberchecked{886} papers. As a result, the amount of change in review scores is quite small.  Figure~\ref{FigDiscussionAbsoluteScores} depicts the score change -- in absolute value -- averaged across all reviewers and all papers. While the allowed range of the scores is 1 to 5, the change in mean score is less than 0.1.  

From the point of view of reviewers, we see a significant correlation between participation in the discussions and the final decisions. Specifically, for each paper we computed the average of scores given by all reviewers who participated in the discussions and the average of scores given by all reviewers who did not participate (when there was at least one reviewer of each type). We discarded this paper if both types of reviewers provided an identical average score. Now, if the participating reviewers gave a higher score than the non-participating reviewers and if the paper was accepted, we counted it as an agreement of the final decision with the participating reviewers. If the participating reviewers gave a lower score than the non-participating reviewers and if the paper was not accepted, then also we count it as an agreement of the final decision with the participating reviewers. Otherwise, we counted the paper as having a disagreement between the final decisions and the participating reviewers. From the data, we observe a statistically significant agreement of the final decisions and the participating reviewers with p = $1.6 \times 10^{-6}$ with \effectsize{0.13}. We continue to observe a statistically significant correlation when this analysis is performed restricted to pool 1 reviewers (p = $7.7 \times 10^{-4}$) or to pool 2 reviewers (p = $1.3 \times 10^{-4}$) alone. Of course, we cannot tell the causality from this correlation, as to whether the discussions actually influenced the decisions or not.

All in all, we observe that only a small fraction of the reviews change scores following the rebuttals. Moreover the magnitude of this change in scores is very small. This observation suggests that this rebuttal process may not be very useful. That said, there are various qualitative aspects that are not accommodated in this quantitative aggregate statistic. First, it may be possible that more reviews changed with respect to the text comments but the reviewers just did not bother to change the scores -- we are unable to check this property as there is no snapshot of the text comments before the rebuttal. Second, there are a reasonable number of discussion posts, however, we do not know what fraction of these posts where reviewers shifted from their earlier opinion. Third, the final decisions are correlated positively with the reviewers who participated in discussions. Taking these factors into account, we think that the present rebuttal system should be put under the microscope regarding its value for the time and effort of such a large number of people. It may also be worth trying alternative systems of recourse for authors, such as a formal appeals process, that help to put more focus on the actual borderline cases.


\begin{Summary}
\keyobservations{\item There is little change in scores post-rebuttal and a moderate amount of discussion.
\item Invited and non-student reviewers participate more in the discussions.
\item Final decisions correlated with scores given by reviewers who participated in discussions, even when stratified by individual pools.
}

\actionitems{\item Force every reviewer to change or confirm their scores after the end of the discussion session.
}

\openquestions{\item How to incentivize reviewer participation in rebuttals/discussions?
\item How to de-bias  reviewers from their initial opinion?
\item Compare the amount of discussion and changes in scores with that in open review processes (particularly, when open reviews are used for conferences of this scale).
\item Compare the efficiency of the rebuttal process with a post-decision appeal procedure to catch only cases that deserve discussion (i.e., possible mistakes).
}
\end{Summary}


\newsubsection{Distribution across subject areas}
\label{SecAreas}

\begin{figure}[t!]
\centering
\noindent\incl{1.0}{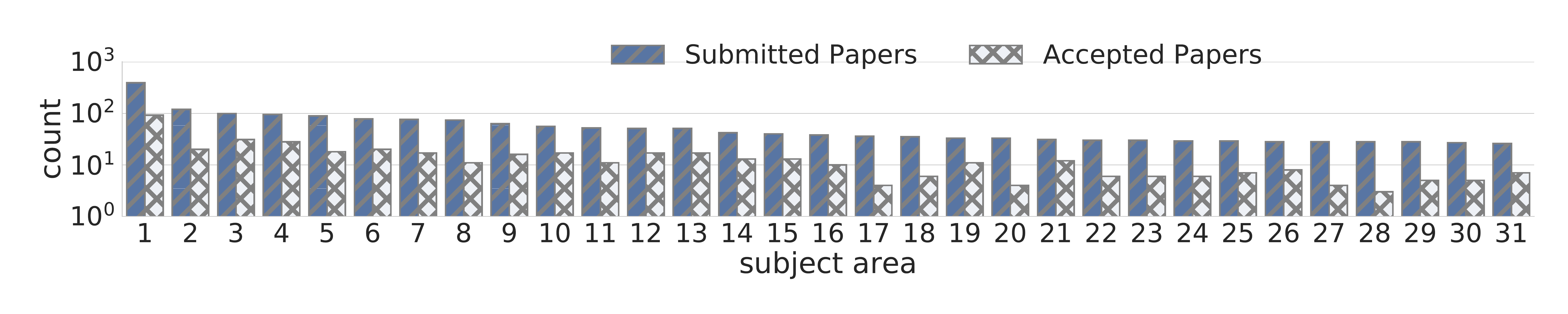} \\\vspace{-.1in}
\noindent\incl{1.0}{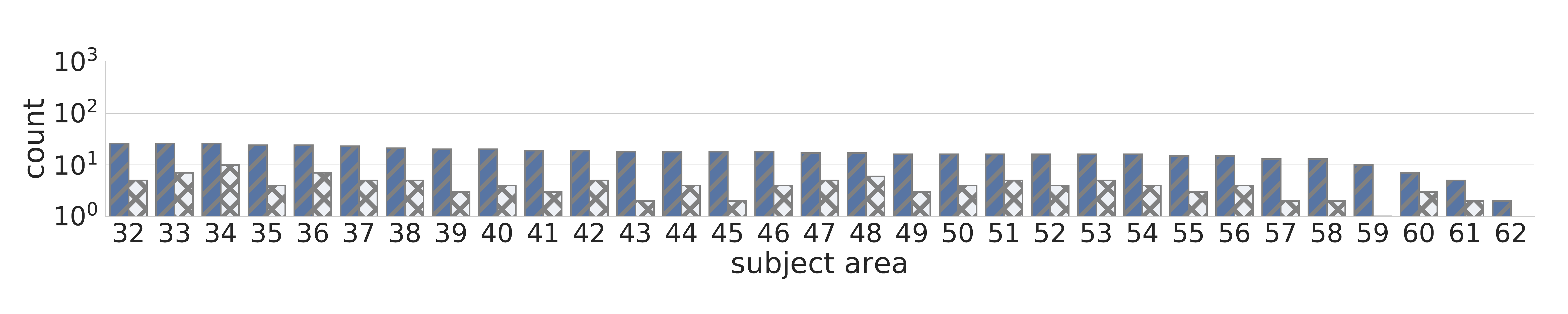}
\caption{The number of submitted and accepted papers per (primary) subject area. The names of the subject areas corresponding to each of the numbers on the x-axis are provided in Appendix~\ref{AppSubjectAreas}.
\label{fig:topics} }
\end{figure}

Figure~\ref{fig:topics}  plots the distribution per subject area (primary subject area), for the submitted papers and for the accepted papers. Of course the proportions are not identical, but the plots do not show any systematic bias either towards  or against any particular areas. The heavy tail in the distributions below also corroborates the significant diversity of topics in the NIPS community. A chi-square test of homogeneity of the two distributions failed to detect any significant difference between the two distributions: \numberchecked{p=$0.6029$}, $\chi^2(dof=62,\#samples=2425)=57.51$.

\begin{Summary}
\keyobservations{\item No observable bias across subject areas in terms of final acceptances}

\actionitems{\item Test for systematic biases for/against any subject area before announcing final decisions}

\openquestions{\item How to assimilate different, subjective opinions of reviewers across different subject areas}
\end{Summary}


\newsubsection{Quantifying the randomness}
\label{SecRandomness}

Quantifying the extent to which the outcome of a peer-review process is different from a random selection of papers is one of the most pressing questions for the  scientific community~\citep{somerville2016bayesian}. 
In this section we conduct two analyses to quantify the randomness in the review scores in NIPS 2016. 


\subsubsection{Messy middle model}
\label{SecMessyMiddle} 
 
\newcommand{\numpapers}{n}
\newcommand{\numagree}{n_{\scalebox{.65}{\mbox{agree}}}}
\newcommand{\numdisagree}{n_{\scalebox{.65}{\mbox{disagree}}}}
\newcommand{\minsamples}{\mu}
\newcommand{\messythreshold}{\alpha}
\newcommand{\fracaccept}{\beta} 
\newcommand{\binoval}{X}
\newcommand{\binovar}{q}
\newcommand{\orderedpapers}{O}
\newcommand{\setmiddlepapers}{M}

The NIPS 2014 experiment~\citep{nips14experiment} led to the proposal of an interesting ``messy middle'' model~\citep{nips14blog}. The messy middle model postulates that the best and the worst papers are clear accepts and clear rejects, respectively, whereas the papers in the middle suffer from random decisions that are independent of the content of the papers. 
The messy middle model is obviously a stylized model, but it nevertheless suggests an interesting investigation into the randomness in the reviews and decisions of the papers that lie in the middle. In this section, we describe such an investigation using the NIPS 2016 data.

The messy middle model assumes random judgments for the middle papers. If the messy middle model were correct then for any pair of papers in the middle, and any pair of common reviewers, the probability of an agreement on the relative ranking of the two papers must be identical to the probability of disagreement. With this model in mind, we restrict attention to the papers in the middle, and then measure how far the agreements of the reviewers are from equiprobable agreements and disagreements. An analysis of this quantity for various notions of the ``middle'' papers yields insight into the messiness in the reviews for papers in the middle.

\noindent {\bf Procedure: }
We now describe the procedure employed for the analysis. Here we let $\numpapers$ denote the total number of papers submitted to the conference and $\fracaccept$ denote the fraction of papers accepted to the conference (we have $\numpapers = 2425$ and $\fracaccept = 0.237$ in NIPS 2016). The procedure is associated to two parameters: $\minsamples$ is the minimum number of samples required and $\messythreshold$ is a threshold of messiness. We choose $\minsamples = 100$ and $\messythreshold = 0.01$ in our subsequent analysis, noting that importantly, our overall conclusions are robust to these choices.

\begin{enumerate}
\item Rank order all papers with respect to their mean scores. Call this ordering as $\orderedpapers$.
\item For every $t \in [0,1]$ and $b \in [0,1]$ (up to some granularity), do the following.\label{EnumStepForloop}
\begin{enumerate}[label*=\arabic*]
\item Initialize variables $\numagree[t,b] = \numdisagree[t,b] = 0$.
\item Consider the set of papers obtained by removing the top $t$ fraction of papers and bottom $b$ fraction of papers from $\orderedpapers$. Call this (unordered) set of ``middle papers'' as $\setmiddlepapers$.
\item If $(\fracaccept - t) \numpapers < \minsamples$ or $((1-\fracaccept)-b)\numpapers < \minsamples$ then continue to the next values of $(t,b)$ in Step~\ref{EnumStepForloop}.\label{EnumMinpapers}
\item 
Consider any pair of reviewers and any pair of papers in $\setmiddlepapers$ that is reviewed by both the reviewers. We say that this pair of reviewers agrees on this pair of papers  if both reviewers provide a higher mean score (taken across the features) to the same paper as compared to the other paper. We say that this pair disagrees if the paper rated higher by one reviewer (in terms of the mean score across the features) is rated lower by the other reviewer. Ties are discarded. We count the total number of such agreements (denoted as $\numagree[t,b]$) and disagreements (denoted as $\numdisagree[t,b]$) within each of the two pools.
\end{enumerate}
\item Find the largest value of $(1\!-\!t\!-\!b)$ such that $(\numagree[t,b]\! +\! \numdisagree[t,b]) \geq \minsamples$ and $\frac{\numagree[t,b]}{\numagree[t,b] + \numdisagree[t,b]} < 0.5 \!+\! \messythreshold$. This largest value of $(1\!-\!t\!-\!b)$ is defined as the size of the messy middle.\label{EnumChooseWindow}
\end{enumerate}

Let us spend a moment interpreting some steps of the procedure. Step~\ref{EnumMinpapers} as well as the $\minsamples$-condition in Step~\ref{EnumChooseWindow} ensures that there are a sufficient number of samples for any computation on the messy middle region. Specifically, the conditions $(\fracaccept - t) \numpapers < \minsamples$ and $((1-\fracaccept)-b)\numpapers < \minsamples$ ensure existence of a sufficient number of papers above and below the acceptance threshold. Under this constraint, Step~\ref{EnumChooseWindow} then finds the largest window of papers in the middle such that the fraction of reviewer-agreements is at most $(0.5+\messythreshold)$. Thus a \emph{smaller} size of the window is a desirable property.

We can now use this analysis to compare messy middle window sizes for two or more conferences. When making such a comparison, we make one adjustment. In the last step (Step~\ref{EnumChooseWindow}), we consider only those values of $(t,b)$ such that $\numagree[t,b] + \numdisagree[t,b] \geq \minsamples$ for both datasets. Then compare the sizes of the messy middle.

\begin{figure}[t!]
\centering
\begingroup
\captionsetup[subfigure]{width=.91\textwidth}
\subfloat[Size of the messy middle windows. 
\label{FigRevMesyMiddle}]{
\begin{tabular}{|c|c|}
\hline
Conference & Size of messy middle\\
\hline 
NIPS 2015 & 45\%\\
NIPS 2016 & 30\%\\
\hline 
\end{tabular}
}
\endgroup\\
\subfloat[Histogram of the variance of acceptance decisions (according to mean scores) of the papers in a bootstrapped analysis.\label{FigBootStrap}]{
\includegraphics[width = 5in]{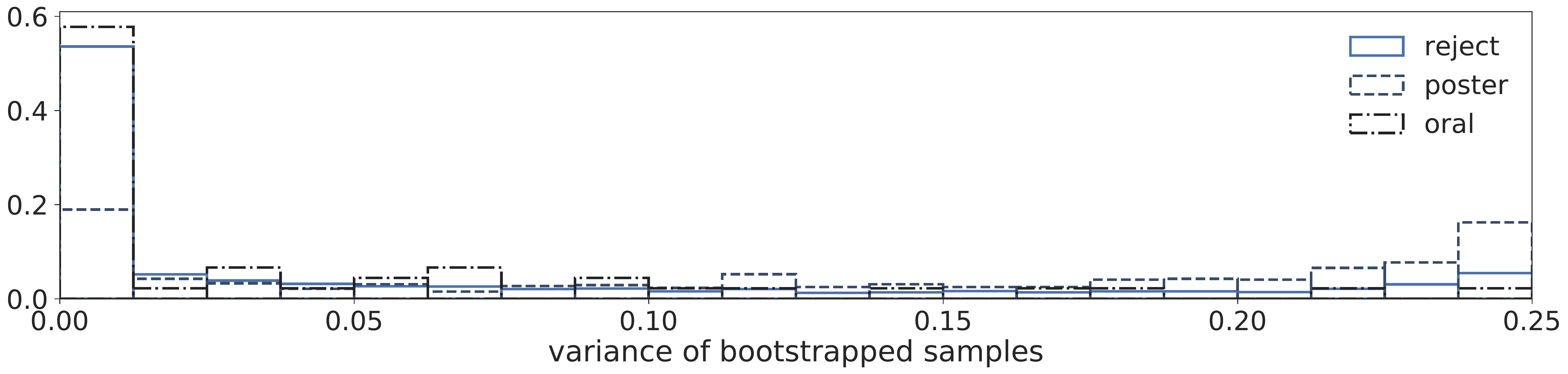}}
\caption{Amount of randomness in the reviews.\label{FigRandomness}}
\end{figure}

\noindent{\bf Results:} 
We used this procedure to compute the size of the messy middle in NIPS 2016 and also in NIPS 2015. The granularity we used is $1/20$, that is, $t,b \in \{0,1/20,2/20,\ldots,1\}$. NIPS 2015 had a marginally higher average number of reviews per paper as compared to NIPS 2016. We set $\minsamples = 100$ and $\messythreshold = 0.01$ (note that the conclusions drawn below are robust to these choices). The results of the analysis are tabulated in Figure~\ref{FigRevMesyMiddle}. 

In the NIPS 2016 data, we observe that the size of the messy middle is $30\%$. Specifically, if we remove the bottom $70\%$ of papers (and none of the top papers) then we see that the inter-reviewer agreements are near-random, but farther from random otherwise. On the other hand, we observe that the size of the messy middle is $45\%$ in the NIPS 2015 data, which occurs when removing $15\%$ of the top papers and $40\%$ of the bottom papers. We thus see that in terms of this metric of the size of the messy middle, the NIPS 2016 review data is an improvement over the previous edition of the conference.

Such an analysis is useful in comparing the noise in the review data across conferences. It can particularly be useful to evaluate the effects of any changes made in the peer-review process. The ease of doing this post hoc analysis, without necessitating any controlled experiment, is a significant benefit to this approach of analysis. In order to enable comparisons of the size of the messy middle of NIPS 2016 with other conferences, we provide the values of $\frac{\numagree[t,b]}{\numagree[t,b] + \numdisagree[t,b]}$ and $(\numagree[t,b] + \numdisagree[t,b])$ for the NIPS 2016 data  for all values of $(t,b)$ in Appendix~\ref{AppMessyValues}.

It is important to note that this post hoc analysis is not strictly comparable to the NIPS 2014 controlled experiment because we do not have access to a true ranking or a counterfactual. That said, since such an analysis can easily be performed post hoc using the data from reviews and does not require any special arrangement in the review process, it would be useful to see how these results compare to the data from other conferences.


\subsubsection{A bootstrapped analysis}

In this section we conduct an analysis to measure the randomness in the reviews in the NIPS 2016 data compared to that of random selection. In our analysis, we first conduct 1000 iterations of the following procedure. For each paper, we consider the set of reviewers who reviewed this paper. We then choose the same number of reviewers uniformly at random with replacement from the set of original reviewers for this paper. We then take the mean of the scores across all features and across all the sampled reviewers for that paper. Next we rank order all papers in terms of these mean scores and choose the top 23.7\% of the papers as ``accepted'' in this iteration and the others as rejected. 

Our analysis focuses on the variance of the acceptance decisions for each paper. At the end of all iterations, for each paper, we compute the fraction of iterations in which the paper was accepted. Letting $\fracaccept_i \in [0,1]$ denote this fraction for any paper $i$, the variance in the acceptance decisions for this paper equals $\fracaccept_i (1 - \fracaccept_i)$. We plot a histogram of the computed variances (for every paper) in Figure~\ref{FigBootStrap}. For comparison, note that in an ideal world, the variance of the decisions for each paper would be zero. Observe that a large fraction of rejected papers as well as a large fraction of papers that were accepted as oral presentations have a near-zero variance. On the other hand, a notable fraction of papers accepted as posters as well as those rejected have a variance close to its largest possible value of $\frac{1}{4}$.

\begin{Summary}
\keyobservations{\item A notable subset of papers incurs ``messy middle'' randomness. 
\item The messy middle region is smaller in NIPS 2016 as compared to NIPS 2015.
\item A bootstrapped analysis shows a significant variance in reviewer scores for a notable fraction of papers that are accepted as posters. A large fraction of papers accepted for oral presentations or rejected have near-zero variance.
}

\actionitems{\item Measure and compare post hoc goodness (using the analysis of this paper or otherwise) of various review processes in order to choose a good review process in a data-dependent manner.
}

\openquestions{\item Principled design of statistical tests for post hoc comparison of goodness of different review processes.}
\end{Summary}

\newsubsection{Ordinal data collection} 
\label{SecOrdinal}

The data collected from the reviewers in the NIPS 2016 review process comprises cardinal ratings (in addition to the free-form text-based reviews) where reviewers score each paper on four features on a scale of 1 to 5. A second form of data collection that is popular in many applications, although not as much in conference reviews, is ordinal or comparative ranked data. The ordinal data collection procedure that we consider asks each reviewer to provide a total ordering of all papers that the reviewer reviewed.

There are various tradeoffs between collecting cardinal ratings and ordinal rankings. In the context of paper reviews, cardinal ratings make reviewers read each individual paper more carefully (and not make snap judgments), and can elicit more than a just one bit of information. On the other hand, ordinal rankings allow for nuanced comparative feedback, help avoid ties, and are free of various biases and calibration issues that otherwise arise in cardinal scores~\citep{harzing2009rating,krosnick1988test,russell1994ranking,rankin1980comparison,cambre2018juxtapeer}. We refer the reader to the papers by~\citet{barnett2003modern, stewart2005absolute,shah2016estimation, shah2016stochastically} and references therein for more details on ordinal data collection and processing. In the present paper, we present three sets of analyses with the ordinal rankings collected from reviewers. 

\subsubsection{Tie breaks}

\begin{figure}[b!]
\centering
\incl{0.4}{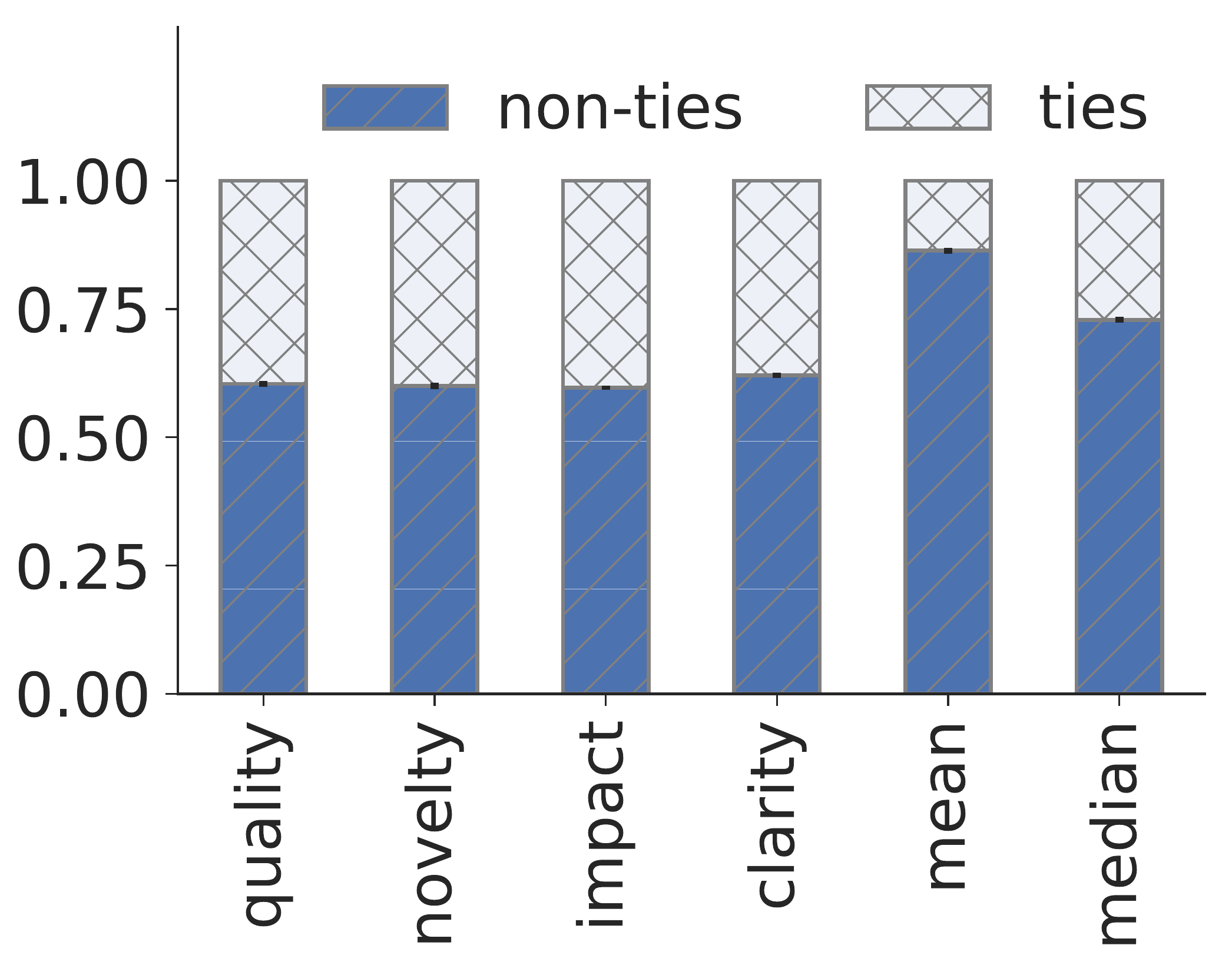}
\caption{Proportion of ties in reviewer scores. The bars titled ``mean'' and ``median'' represent the mean and median scores across all four features.  \label{figure-Ties}}
\end{figure}

\label{SecRanking_tiebreaks}
An ordinal ranking of the papers provided by a reviewer ensures that there are no ties in the reviewer's evaluations. On the other hand, asking cardinal scores can result in scores that are tied, thereby preventing an opportunity for the AC to discern a difference between the two papers from the provided scores.

\begin{figure}[b]
\centering

\subfloat[\label{figure-OrdInterWorkerAgree}]{\includegraphics[height = 1.2in]{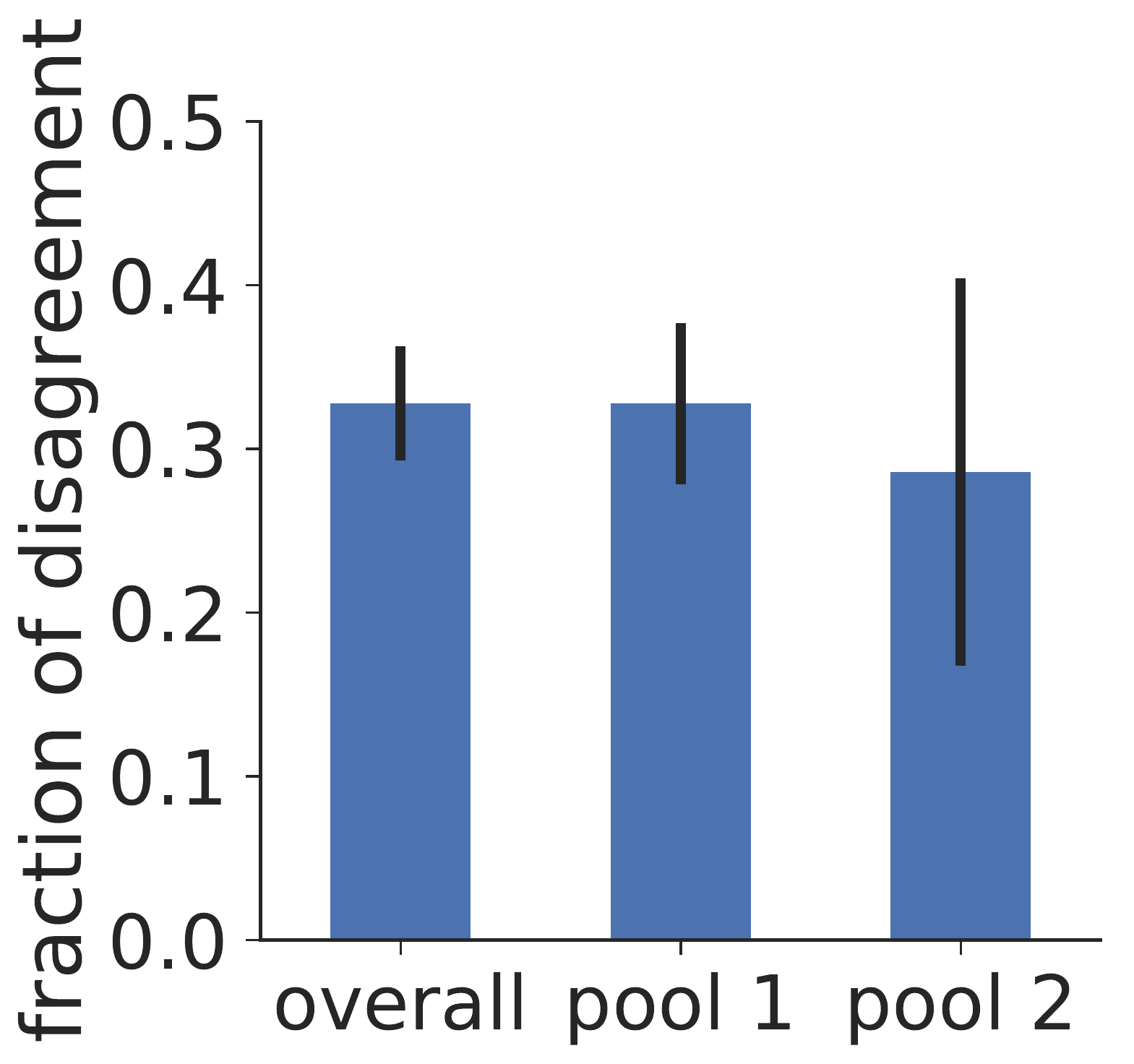}}~~
\subfloat[\label{FigureOrdinalConsistency}]{\includegraphics[height = 1.2in]{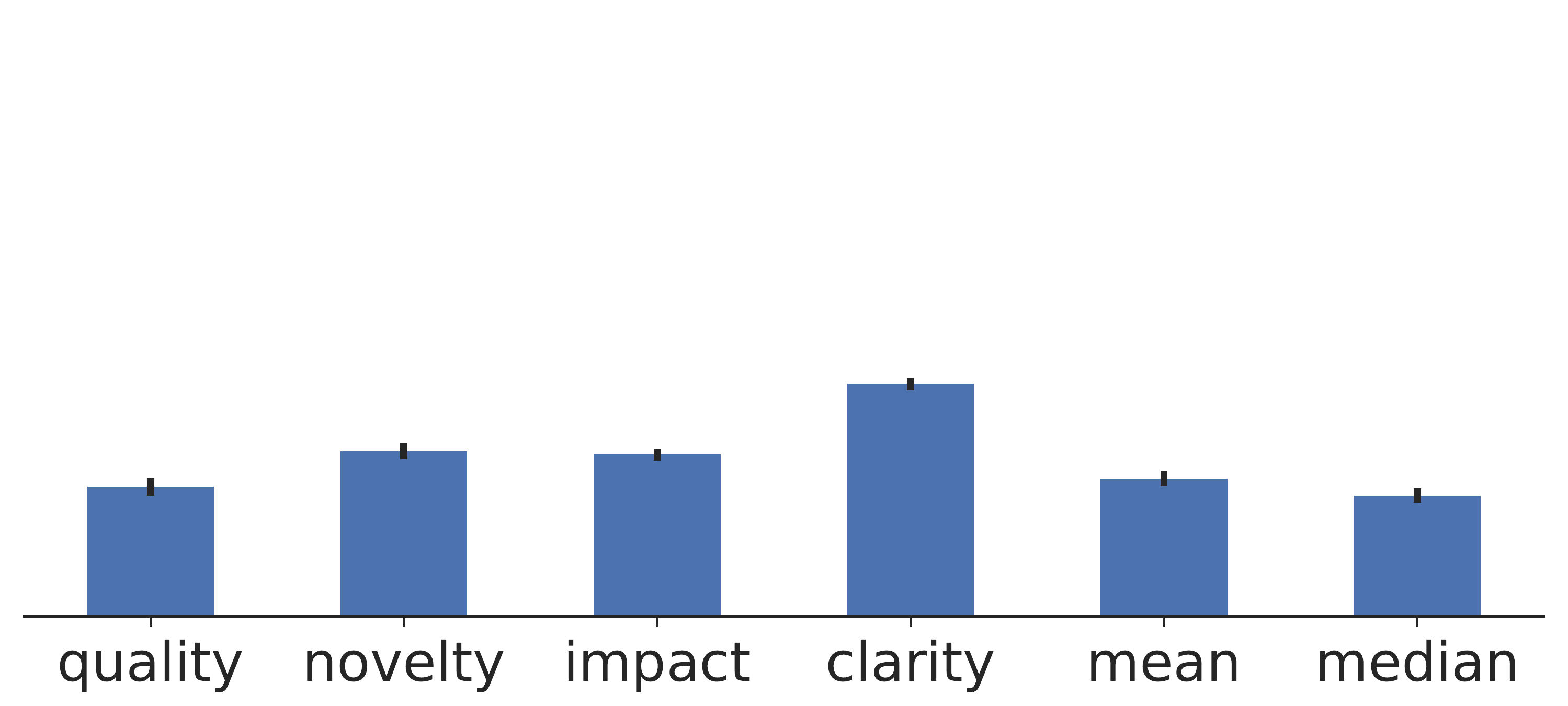}}~~
\subfloat[\label{figure-OrdinalDecisions}]{\includegraphics[height = 1.2in]{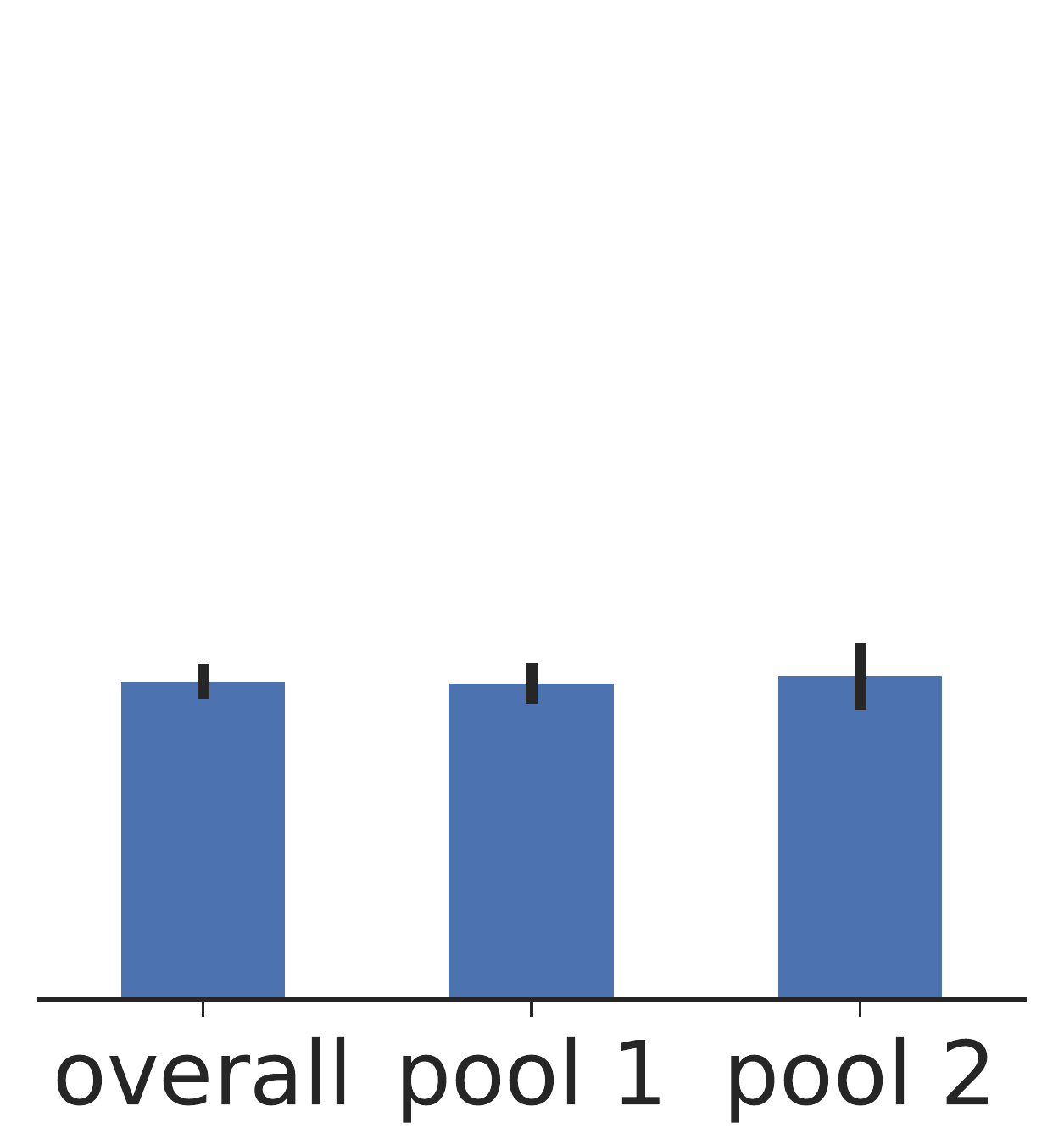}}
\caption{Fraction of disagreements (a) within ordinal rankings between different pairs of reviewer types; (b) between ordinal rankings and cardinal ratings (``mean'' and ``median'' refer to the mean and median of the cardinal scores for the four features); and (c) between ordinal rankings and final acceptance decisions.}
\end{figure}

In order to evaluate the prevalence of ties under cardinal scores, we performed the following computation. For every \tuple{\texttt{paper}, \texttt{paper}, \texttt{reviewer}} triplet such that the reviewer reviewed both papers, and for any chosen feature (i.e., quality, novelty, impact, and clarity), we computed whether the reviewer provided the same score to both papers or not. We totaled such ties and non-ties across all such triplets. 

Figure~\ref{figure-Ties} depicts the proportion of ties computed across all submitted papers. The total sample size is \numberchecked{26106}. Observe that a significant fraction -- exceeding 30\% for each of the four features -- of pairs of reviewer scores are tied. When only the \toptwok papers were used in the calculation, the fraction of ties in each feature is even higher,  by approximately $10\%-15\%$ of the respective value in the setting of all papers. In conclusion, these results reveal a significant proportion of ties in the cardinal scoring scheme and also confirm that, by design, ties are inevitable in this scoring scheme. The use of ordinal rankings, on the other hand, does not suffer from such a drawback.

\subsubsection{Consistency of ordinal ranking data}
\label{SecRanking_consistency}

While there is substantial literature on benefits of collecting data in an ordinal ranking form, several past works also recommend verifying if the application setting under consideration is appropriate for ordinal rankings. 
For instance,~\citet{russell1994ranking} state the benefits of ranking for settings ``where the items are highly discriminable'';~\citet{peng1997validity} ask respondents to rank 18 values in order of importance but observe unstable and inconsistent results;~\citet{harzing2009rating} argue that ranking generally requires a higher level of attention than rating and that asking respondents to rank more than a handful of statements puts a
very high demand on their cognitive abilities. 
Accordingly, this section is devoted to performing sanity checks on the ordinal ranking data obtained subsequent to the NIPS 2016 review process. We do so by comparing certain measures of consistency of the ordinal data with the cardinal data obtained in the main review process.


\xhdr{Agreements within ordinal rankings}
For every pair of papers that have two reviewers in
common, we compute whether these two reviewers agree on the relative ordinal ranking
of the two papers or if they disagree. In more detail, we say that this pair of reviewers agrees on this pair of papers if both reviewers rank the same paper higher than the other in their respective ordinal rankings; we say that this pair disagrees if the paper ranked higher by one reviewer is rated lower by the other. Figure~\ref{figure-OrdInterWorkerAgree} depicts the proportion of disagreements for the ordinal rankings in the entire set of papers, as well as broken down by the type of reviewer. First, observe that the ordinal rankings have a similar level of consistency as that observed in the cardinal scores in Figure~\ref{FigureCardInterWorkerAgreeAfter}. Second, we observe no statistically significant difference between the two pools: \numberchecked{p=$0.9849$} for Pearson's chi-squared test and effect size \numberchecked{\effectsize{0.0018}}. The sample sizes are \numberchecked{696, 348 and 56} for all reviewers, pool 1 and pool 2 respectively.

\xhdr{Agreement of ordinal rankings with cardinal ratings} Let us now evaluate how well the overall ordinal rankings associate with the cardinal scores given for the individual features.  
For every pair of
papers that have a common reviewer, we compare whether the relative ordering of the
cardinal scores for a given feature agree with the ordinal ranking given by the
reviewer for the pair of papers. We report the proportion of disagreements in Figure~\ref{FigureOrdinalConsistency}.  We observe the high amount of agreement of the ordinal rankings with the cardinal scores -- for instance, the median cardinal score agrees in about 90\% of cases with the overall ordinal rankings provided by the reviewers. 

\xhdr{Agreement of ordinal rankings with final decisions}
We finally compute the amount of agreement between the ordinal rankings provided by the reviewers and the final decisions of acceptance. We consider all \tuple{paper, paper, reviewer} triplets where the reviewer reviewed both papers, and one of these papers was eventually accepted and the other was rejected. For every such triplet, we evaluate whether the reviewer had ranked the accepted paper higher than the rejected paper (``agreement'') or vice versa (``disagreement''). We report the proportion of agreements and disagreements in Figure~\ref{figure-OrdinalDecisions}. We see that the agreement of the overall rankings with the eventual decisions is quite high -- there are roughly five agreements for every disagreement. 

When restricted to the \toptwok papers, we observe that the disagreements of ordinal rankings with final decisions increase to 27-28\% in all three categories (overall, pool 1 and pool 2) from 16-17\% in the case of all papers. 
Note that the experiments on inter-reviewer agreements do not permit an effective analysis when restricted to \toptwok papers as the sample size reduces quadratically (that is, reduces to a fraction $.47^2 \approx .2$ of the sample size with all papers).


\subsubsection{Detecting anomalies}

Ordinal rankings can be used to detect anomalies in reviews. We discuss this aspect in the Section~\ref{SecAnomalies}.

\begin{Summary}
\keyobservations{

\item Ordinal rankings are a viable option for collecting reviewer opinions.

\item There are a large number of ties in ratings provided by reviewers: there are more than 30\% ties in each feature and even greater fraction of ties in the \toptwok papers.

\item Ordinal rankings can be used to check inconsistencies in the reviews.
}

\actionitems{\item Use a hybrid collection method with cardinal ratings for individual features and an overall ordinal ranking to avail benefits of both data-collection methods.
}

\openquestions{\item Perform controlled experiments in order to quantify the benefits and possible issues with ordinal rankings. 

\item Design algorithms to efficiently combine cardinal ratings for features and ordinal overall rankings to provide useful guidelines to area chairs for their decisions.}

\end{Summary}


\newsubsection{Checking inconsistencies}
\label{SecAnomalies}

In this section, we propose an automated technique to help reduce some human errors and inconsistencies in the review process. In particular, we propose to automatically check for inconsistencies in the review ratings provided by the reviewers. On finding any such inconsistency, we propose to then have the area chairs either manually investigate this inconsistency or to manually or automatically contact the reviewer requesting an explanation. In what follows, we propose two notions of inconsistencies in regards to the NIPS 2016 review process and quantify their presence in the NIPS 2016 review data.

\xhdr{Anomalies in feature ratings} We investigate whether any reviewer indicated that paper ``A'' is strictly better than paper ``B'' in all four features, but rank paper ``A'' lower than paper ``B'' in the ordinal ranking. We find that there are \numberchecked{55} such pairs of reviews provided by \numberchecked{44} distinct reviewers. If we restrict attention to the \toptwok papers, we find that that there are \numberchecked{10} such pairs of reviews provided by \numberchecked{10} distinct reviewers.\footnote{Note that the total number of pairs of papers reduces more than 4-fold when moving from the set of all papers to the \toptwok set.} 

\xhdr{Anomalies in fatal flaws} We now investigate if there are cases when a reviewer indicated a fatal flaw in a paper, but that reviewer ranked it above another paper that did not have a fatal flaw according to the reviewer. We found \numberchecked{349} such cases across \numberchecked{176} such reviewers. 
The proportion of such cases is similar among volunteer and invited reviewers. Among the \toptwok papers, there are \numberchecked{55} such pairs across \numberchecked{33} reviewers.

One may think that the number of such cases is large because ordinal survey was done after the review process, so people may not have remembered the papers well or may not have done a thorough job as they knew it would not count towards the reviews. However, the ordinal data actually is quite consistent with the cardinal data (Section~\ref{SecRanking_consistency}). Hence we do not think such a large discrepancy with fatal flaws can be explained solely due to such a delay-related noise.

Two possible explanations for such anomalies are as follows. Either the reviewer may not have done an adequate job of the review, or the set of provided features are grossly inadequate to express reviewers' opinions. In either case, we suggest automatically checking for such glaring inconsistencies (irrespective of whether ordinal or cardinal final ratings are used) during the review process, and contacting the respective reviewers to understand their reasoning.\footnote{This analysis was performed after completion of the review process, and hence reviewers were not contacted for these inconsistencies.} We hope that such a checkpoint will be useful in improving the overall quality of the review process.

\begin{Summary}
\keyobservations{\item 55 instances (across 44 reviewers) of a reviewer rating a paper higher than another for all features but inverting the relative ranking of the two papers in the overall ordering.

\item 349 cases where a reviewer indicated a fatal flaw in a paper but ranked it higher than another paper without any indicated fatal flaw.}

\actionitems{\item Check for inconsistencies in the reviews and contact respective reviewers.
}

\openquestions{\item What are other inconsistencies that can be checked in an automated manner?}

\end{Summary}


\section{Discussion and conclusions}
\label{SecConclusion}

NIPS has historically been the terrain of much experimentation to improve the review process and this paper is our contribution to advance the state of the art in review process design. In this paper, we reported a post hoc analysis of the NIPS 2016 review process. Our analysis yielded useful insights into the peer-review process, suggested action items for future conferences, and resulted in several open problems towards improving the academic peer-review process, as enumerated throughout this paper. 

Our tools include several means of detecting potential artifacts or biases, and statistical tests to validate hypotheses made: Comparing the distribution of topics in submitted papers and accepted papers; creating a graph of proximity of reviewers (according to commonly reviewed papers) and papers (according to common reviewers) to detect potential disconnected communities; test to compare two pools of reviewers; quantifying the noise in the review scores. We also observed that the histogram of scores obtained included a significantly larger fraction of papers than the guidelines suggested. This observation suggests a more careful design of the elicitation interface and the type of feedback provided to authors.

Selection biases that arise when recruiting reviewers and ACs in a review process of this scale are difficult to deal with. Some designs in the selection of reviewers lend themselves more to bias than others. In NIPS2016, we made some design choices of the review process with the intention of reducing these biases. For instance, the recruitment of volunteer author-reviewers helped increase the diversity of the reviewer pool. They were less prone to selection bias compared to selecting reviewers by invitation only, primarily based on AC recommendations. With respect to reducing bias across AC decisions, we introduced the ``AC buddy system'' in which pairs of ACs had to make decisions jointly about all their papers. This method scales well with the increase in number of papers, but is sub-optimal to calibrate well decisions since buddy pairs form disjoint decision units (no paper overlap between buddy pairs). However, decision processes based on a conference between several or all ACs, as done in earlier editions of the conference, are also not perfect because decisions are sometimes dominated by self-confident and/or opinionated ACs. Although the evidence we gathered from our analyses did not reveal any ``obvious'' bias, it does not mean that there is none. We hope that some designs of our review process will shed some lights on ways of improving bias-immune or bias-avoidance procedures for future conferences.

The reviews themselves were of mixed quality, but recruiting more reviewers (between 4 and 6 per paper) ensured that each paper had a better chance to get a few competent reviews. We gave a strong role to the ACs who arbitrated between good and bad reviews and made the final decision, which was not just based on an average score. To recruit more reviewers (and possibly a more diverse and less biased set of reviewers) we introduced the new idea to invite {\em volunteer} author reviewers, which we think is a good contribution. In  particular,  next  to many  PhD  students,  this  brought  a  considerable  amount  of  senior  reviewers  in  the  system  as well. Some  of  the  ACs  systematically disregarded  volunteer  reviews, judging that they could not be trusted. But, our analysis did not reveal that reviewers from that pool made decisions significantly different from the pool of reviewers invited by recommendation. However, more senior reviewers seem to put more effort into providing detailed reviews, and participating to rebuttals and discussions. Hence we need to find means of encouraging and possibly educating more junior reviewers to participate in these aspects. As a means of self-assessment and encouragement, reviewers could receive statistics about review length, amount of agreement between reviewers, and participation to rebuttals and discussions, as well as figures concerning their own participation. Naturally, the participation of junior reviewers in the review process is a form of education. It would be nice to track from year to year whether individual reviewers ramp up their review length, level of agreement with other reviewers, and participation in discussions and rebuttals. Note that we believe that such statistics should not be used as a means of selecting reviewers because this could bias the selection.

It is an on-going debate to which extent the decision process should be automated and what means could be used to automate it. We provide some elements to fuel this discussion. We evaluated how rebuttals and discussions change the scores. Although this concerns only a minority of papers, we believe that ACs have a key role in arbitrating decisions when there is a controversy and that this is not easy to monitor merely with scores. Since scores do not seem to be consistently updated by reviewers after rebuttal/discussions, maybe the review process should include a score confirmation to make sure that absence of change in score is not due to negligence. Mixing ordinal and cardinal scores may reduce the problems of reviewer calibration, tie breaking, and identifying anomalies possibly due to human error.

All in all, it is important to realize that in a review process of this scale, there is not a single person who really controls what is going on at all levels. Program chairs spend a lot of time on quality control, but definitely cannot control the decisions on all individual papers or the quality of individual reviewers. In the end, we have to trust the area chairs and reviewers: the better reviews \emph{all of us} provide, the better the outcome of the review process. We as a community must also continue to strive improving the peer-review process itself,  via experiments, analysis, and open discussions. This topic in itself is a fertile ground for future research with many useful open problems including those enumerated throughout the paper.

\subsection*{Acknowledgments}
This work would not have been possible without the support of the NIPS foundation and the entire committee of the NIPS 2016 conference. We thank the program chairs and program managers of the NIPS 2015 conference for their help during the organization. We thank Baiyu Chen for preliminary experiments conducted on ordinal data. Isabelle Guyon acknowledges funding from the Paris-Saclay scientific foundation. Ulrike von Luxburg has been supported by the Deutsche Forschungsgemeinschaft (Institutional Strategy of the University of Tübingen, ZUK 63). Krikamol Muandet acknowledges fundings from the Faculty of Science, Mahidol University and the Thailand Research Fund (TRF).

\bibliographystyle{plainnat}
\setlength{\bibsep}{7pt}
\bibliography{references}

\newpage
\appendix

\noindent{\bf \Large APPENDIX}\\

In the appendix we present some additional details about the experiments. 

\section{Subject areas}
\label{AppSubjectAreas}

Here are the subject areas associated to the subject area numbers in Figure~\ref{fig:topics}. 

\setlength{\tabcolsep}{0.02em}
\renewcommand{\arraystretch}{1.0}
\begin{tabular}{>{\footnotesize}l>{\footnotesize}l>{\footnotesize}l}
1. Deep learning/Neural networks && 32 Causality\\2. (Application) Computer Vision&& 33 Bayesian nonparametrics \\3. Learning theory && 34 Variational inference \\4. Convex opt. and big data && 35 Similarity and Distance Learning\\5. Sparsity and feature selection&& 36 (Other) Statistics\\6. Clustering && 37 Spectral methods \\7. Reinforcement learning && 38 Active Learning  \\8. Large scale learning  && 39 Graph-based Learning\\9. Graphical models&& 40 (Other) Bayesian Inference\\10. Bandit algorithms&& 41 (Application) Collab. Filtering / Recommender Systems\\11. Matrix factorization && 42 Information Theory \\12. Online learning && 43 (Application) Signal and Speech Processing	\\13. (Other) Optimization&& 44 (Application) Social Networks\\14. (Other) Neuroscience&& 45 (Other) Robotics and Control \\15. Kernel methods&& 46 Nonlin. dim. reduction \\16. Gaussian process && 47 Model selection and structure learning \\17. Multitask/Transfer learning&& 48 Ensemble methods and Boosting \\18. Component Analysis (ICA, PCA, \ldots) && 49 Stochastic methods\\19. Combinatorial optimization&& 50 (Other) Cognitive Science\\20. Time series analysis&& 51 Structured prediction \\21. (Other) Probabilistic Models and Methods&& 52 Ranking and Preference Learning\\22. (Other) Applications&& 53 Game Theory and Econometrics \\23. (Other) Machine Learning Topics&& 54 (Application) Privacy, Anonymity, Security	\\24. (Cognitive/Neuro) Theoretical Neuroscience&& 55 (Cognitive/Neuro) Perception\\25. (Other) Unsupervised Learning Methods&& 56 (Application) Bioinfo. and Systems Bio.\\26. MCMC&& 57 Regularization and Large Margin Methods\\27. Semi-supervised && 58 (Other) Regression\\28. (Other) Classification&& 59 (Application) Information Retrieval\\29. (Application) Natural Language and Text	&& 60 	(Application) Web App. and Internet\\30. (Application) Object and Pattern Recognition	&\quad& 61 	(Cognitive/Neuro) Reinforcement Learning\\31. (Cognitive/Neuro) Neural Coding	&& 62 (Cognitive/Neuro) Language
\end{tabular}

\section{Messy middle details}
\label{AppMessyValues}

\begin{figure}[b!]
\includegraphics[width=\textwidth]{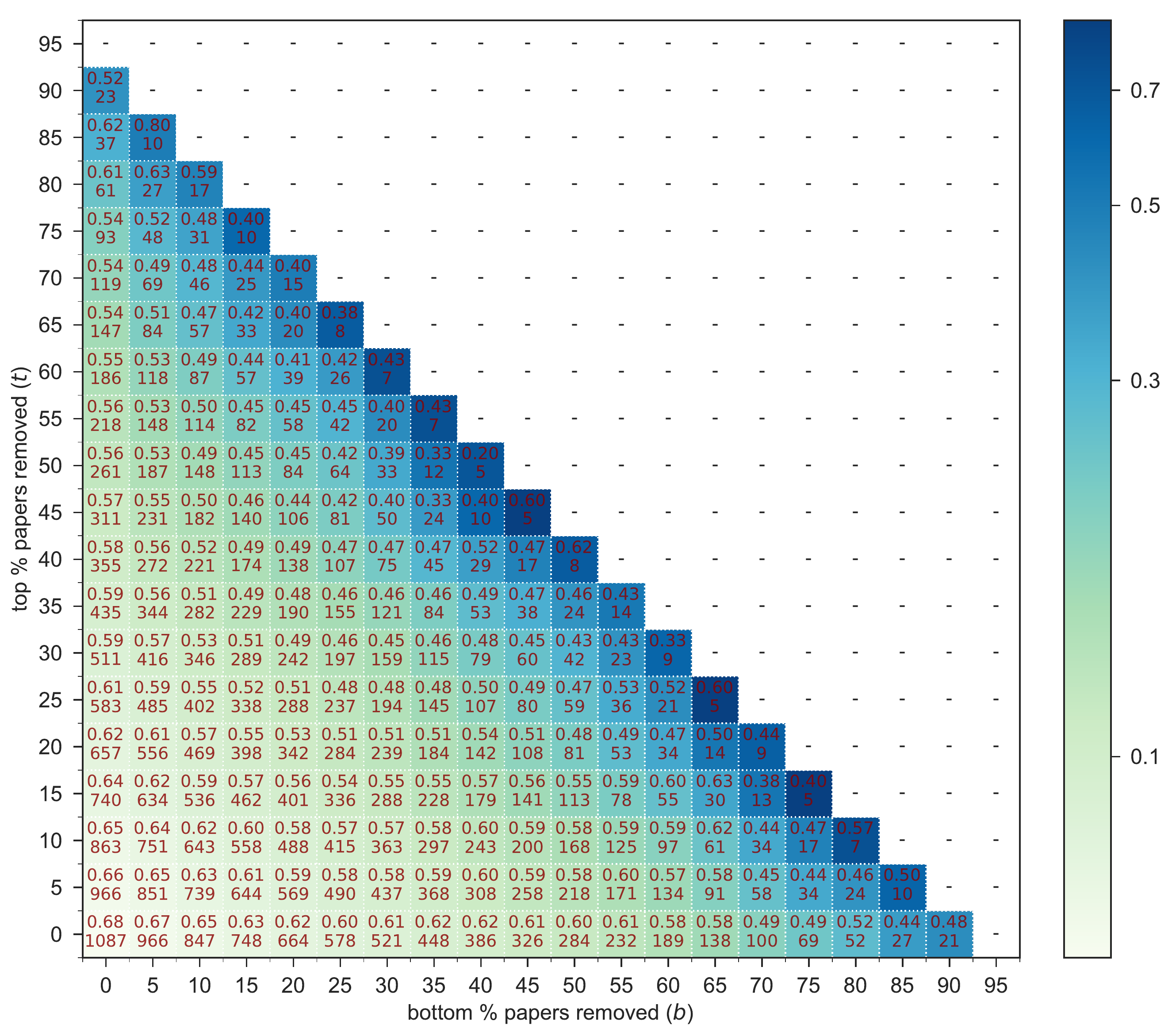}
\caption{The inter-reviewer agreement ratios in the messy middle. For each value of $t$ and $b$, we report two numbers: The agreement ratio $r := n_{\text{agree}} / (n_{\text{agree}}+n_{\text{disagree}})$ and the number of overlapping paper-reviewer pairs $m := n_{\text{agree}} + n_{\text{disagree}}$. Each cell is color-coded by the size of the 95\% confidence interval (on a log scale).
\label{ConfInvTable}}
\end{figure}

In Figure~\ref{ConfInvTable} we provide the values of the fraction of agreements $r := \frac{\numagree[t,b]}{\numagree[t,b]+\numdisagree[t,b]}$ at the top of the corresponding cell and number of pairs $m:=(\numagree[t,b]+\numdisagree[t,b])$ for every value of $(t,b)$ at the bottom of the corresponding cell. Note that the values are computed for all values of $(t,b)$ ignoring the sample size restriction imposed by Step~\ref{EnumMinpapers} of the procedure outlined in Section~\ref{SecMessyMiddle}. Each cell in the table is color-coded by the size of the 95\% confidence interval (on a log-scale) computed as $(2 \times 1.96)\sqrt{ \frac{r (1-r)}{m}}$. 

\end{document}